\documentclass[aps, prd, twocolumn, showpacs, floatfix, letterpaper,
  nofootinbib, superscriptaddress,longbibliography]{revtex4-2}

\usepackage{amsmath, amssymb, amsfonts, physics, bm}
\usepackage{graphicx}
\usepackage{epsfig}
\usepackage{pgf}
\usepackage{color}
\usepackage[T1]{fontenc}
\usepackage[utf8]{inputenc}
\usepackage{graphicx}
\usepackage[english]{babel}
\usepackage{makecell}
\usepackage{hyperref}
\usepackage[draft,deletedmarkup=xout]{changes}
\usepackage{mathtools}
\usepackage{soul}
\usepackage{ulem}
\usepackage{relsize}
\usepackage{orcidlink} 
\usepackage{bbm}
\usepackage{bbold}
\usepackage{newunicodechar}
\newunicodechar{̂}{\^{} }
\usepackage{mathrsfs}
\usepackage{tensor}

\definecolor{green}{rgb}{0.19,0.64,0.54}
\definecolor{blue}{rgb}{0,0,1}
\definecolor{reddish}{rgb}{0.65, 0.2,0.2}
\definecolor{darkgreen}{rgb}{0.2,0.7,0.3}
\definecolor{darkblue}{rgb}{0.3,0.40,0.48}
\definecolor{gray}{rgb}{.8,.8,.8}

\hypersetup{,
  pdftitle={Biverse},
  pdfauthor={K. Mazde, L. Mickel  and P. Peter},
  colorlinks=true, 
  linkcolor=darkblue, 
  citecolor=darkgreen, 
  filecolor=reddish, 
  linktocpage=true
  }

\newcommand{\Rez}{\real\mathrm{e}}
\newcommand{\Imz}{\imaginary\mathrm{m}}

\newcommand{\Hu}{\mathcal{H}}

\newcommand{\mS}{\mathcal{S}}
\newcommand{\mP}{\mathcal{P}}

\newcommand{\mH}{\mathcal{H}}
\newcommand{\pd}{\partial}

\newcommand{\ii}{\mathsf{i}}

\begin{document}

\title{Quantum cosmological background superposition and
perturbation predictions}

\author{Kratika Mazde\,\orcidlink{0000-0002-6232-4743}}
\email{kratika.mazde@iap.fr}
\affiliation{Institut d'Astrophysique de Paris, CNRS and Sorbonne 
Universit\'e, UMR 7095 98 bis boulevard Arago, 75014, Paris, France}

\author{Lisa Mickel\,\orcidlink{0000-0002-6749-8494}}
\email{lisa.mickel@iap.fr}
\affiliation{Institut d'Astrophysique de Paris, CNRS and Sorbonne 
Universit\'e, UMR 7095 98 bis boulevard Arago, 75014, Paris, France}

\author{Patrick Peter\,\orcidlink{0000-0002-7136-8326}} 
\email{patrick.peter@iap.fr}
\affiliation{Institut d'Astrophysique de Paris, CNRS and Sorbonne 
Universit\'e, UMR 7095 98 bis boulevard Arago, 75014, Paris, France}

\begin{abstract}
Predictions from early universe cosmology typically concern primordial
perturbations generated during epochs where effects arising from the
quantum nature of gravity may be important; quantum vacuum
fluctuations being stretched to cosmological scales during a phase of
inflation. Quantizing the background is then done by assuming a single
close-to-classical state over which perturbations grow, as well as a
Born-Oppenheimer factorization throughout the relevant phase. We
present a scenario in which although the latter factorization remains
valid at all times, we allow the background state to be very
non-classical by defining quantum trajectories through an eikonal
approximation.  We find that these trajectories asymptotically
reproduce an almost classical behavior for the background, but the
predictions for the power spectrum of perturbations can significantly
differ.
\end{abstract}

\maketitle

\section{Introduction}

General Relativity (GR) predicts the existence of regions of spacetime
with diverging energy densities and curvature, namely
singularities. These can be hidden inside black holes or represent
the starting point of Big-Bang cosmology~\cite{Peter:2013avv},
the latter of which we are concerned
with here.  Assuming the cosmic censorship
hypothesis~\cite{VandeMoortel:2025ngd} holds, black hole singularities
can be contained behind an event horizon; the cosmological one,
however, lies at the origin of the Universe and demands an
explanation.

The seemingly simplest option that has been suggested to resolve the
primordial singularity consists in modifying the underlying model on
which cosmology is based, which is an isotropic and homogeneous
solution of GR sourced by a minimally coupled perfect
fluid~\cite{Peter:2013avv}.  One is thus naturally led to consider
either a non Friedman-Lema\^{\i}tre-Robertson-Walker (FLRW) isotropic
and homogeneous metric~\cite{Ganguly:2021pke} or to add additional
degrees of freedom in the form of modifying the gravitational theory
or the matter content~\cite{Cai:2013vm, Battefeld:2014uga}.  Finally,
one can couple matter to gravity in a non-minimal
way~\cite{Granda:2010hb}.  All these options rely on the classical
behavior of the background Universe with superimposed quantum
perturbations generating the large scale structures; they replace the
singularity by a minimal scale factor (maximal curvature), i.e. a
bounce~\cite{Brandenberger:2016vhg}.

In view of the many issues (e.g., instabilities) shared by such
models~\cite{Battefeld:2014uga}, classical extensions of GR may not be
the best way to avoid the cosmological singularity.  The problem has
various similarities with the so-called ultraviolet catastrophe, and
it has been proposed that a quantum version of gravity, initially in
the framework of string theory, might provide a
solution~\cite{Gasperini:1992em}. Since then, many models have been
suggested within various quantum gravity
approaches~\cite{ashtekar2006quantum,
bojowald2001absence,Pinto-Neto:2013toa,Kiefer:2025udf}.

In what follows, we consider a model based on canonical quantization
in the reduced phase space of cosmology within the Arnowitt, Deser,
and Misner (ADM) 3+1 formalism~\cite{arnowitt2008republication} for a
flat FLRW spacetime. Such minisuperspace models are widely used in the
literature (see, e.g. Ref.~\cite{Kiefer:2025udf} and references
therein), as the straightforward quantization of the reduced phase
space {\sl \`a la} Wheeler-DeWitt (WDW) leads to a well-defined
quantum description which is expected to be valid in the regime in
which quantum gravity corrections become important while remaining
sufficiently below the Planck scale.

The WDW equation is, however, still riddled with problems, including
that of time~\cite{kuchar1991problem,isham1993canonical}, which stems
from the fact that the quantized Hamiltonian results in a timeless
Schr\"odinger equation.  There are various ways to deal with this
issue~\cite{Anderson:2012vk,Kiefer:2021zdq}, and we shall take the
view that an internal degree of freedom evolving monotonously, in our
case a perfect fluid, can serve as a clock.  This is also the
framework assumed in Ref.~\cite{Bergeron:2023zzo}, which is based on
an affine quantization procedure~\cite{Klauder:2003ih,Klauder:2012uf,
Bergeron:2013ika} and introduces well-defined semi-classical
(coherent) states for an FLRW universe.  Using the affine quantization
procedure leads to a bouncing universe, and the results of
Ref.~\cite{Bergeron:2023zzo} serve as a starting point for our work.

In general, quantum cosmological models~\cite{Hartle:1983ai,
Vilenkin:1983xq, Halliwell:1984eu} rely on two (implicit) assumptions,
the first being that the background state $\ket{\psi_\textsc{b}(a)}$
ought to be a sufficiently peaked wave function of the scale factor
$a$ so that the expectation value $\expval{a(t)}$ can be used to
provide an effective background evolution and can be used as a source
for the perturbations. The second assumption is that the full state
$\ket{\Psi}$ can be factorized in a "Born-Oppenheimer" way as the
tensor product of the background and perturbation state $\ket{\Psi}
= \ket{\psi_\textsc{b}(a)}
\otimes \ket{\psi_\textsc{p}(\delta g_{\mu\nu},a)}$, with the
$\delta g_{\mu\nu}$ the perturbation degrees of freedom for the
overall metric over the FLRW background (see, for instance, 
Refs.~\cite{Pinho:2006ym, Peter_2006, Peter:2008qz, Lehners:2008vx,
Malkiewicz:2020fvy, Pinto-Neto:2013toa, Ashtekar:2009mb,
Gomar:2014faa, Kamenshchik:2013msa, Chataignier:2020fap,
Bini:2013fea, Kamenshchik:2020yvs}). The latter has been
discussed at length in Refs.~\cite{Bergeron:2024art,
Bergeron:2025eda} and we shall, in what follows, focus on
the former. 

Specifically, we consider a superposition of semi-classical states and
instead of projecting onto one of these states to determine the value
of the scale factor in the early Universe, we make use of the
trajectory approach to quantum mechanics, which gives a time-dependent
function for the scale factor in a natural
way~\cite{Pinto-Neto:2013toa,Pinto-Neto:2018zvn}.  The trajectories
stemming from superposed states exhibit additional, highly
non-classical features in the scale factor evolution.  The next
natural question is then if and how such non-classical features could
manifest observationally and we therefore examine their impact on the
evolution of tensor perturbations.

This paper is organized as follows: we first introduce, in
Sec.~\ref{sec:classical}, the classical ADM Hamiltonian for a flat
FLRW spacetime with dynamics driven by a perfect fluid.  We quantize
the Hamiltonian according to the affine coherent state based procedure
in Sec.~\ref{sec:quantisation}, mostly recalling the results of
Ref.~\cite{Bergeron:2023zzo}, before discussing the possible choices
and conditions for a physically acceptable background state in
section~\ref{sec:bckgS}. Having chosen and parametrized an explicit
two-state situation (biverse), we discuss a natural way of providing
meaningful trajectories in Sec.~\ref{sec:dBB}, whose properties we
investigate through a numerical evaluation in Sec.~\ref{sec:num}.
These new phase-space trajectories lead to potentially observable
effects as shown by the example of tensor perturbations, which we
examine in section~\ref{sec:perturbations}. In the final
section~\ref{sec:conc}, we present our conclusions.

\section{Classical Model}
\label{sec:classical}

We first outline the classical model of the spacetime and its matter
content before proceeding to describe the quantization procedure in
the next section. We work in a flat FLRW spacetime, described by the
metric
\begin{align}
\mathrm{d}s^2 = -N^2(T)\dd T^2 + \underbrace{a^2(T) \delta_{ij}}_{\gamma_{ij}(T)}
\dd x^i\dd x^j\,,
\end{align}
filled with a perfect fluid (we use units for which both the velocity
of light and the reduced Planck constant $\hbar$ are unity).  The
action of the total system is
described by the sum $\mS = \mS_{\textsc{eh}} + \mS_\text{fluid}$ of
the Einstein-Hilbert action
\begin{align}
\mathcal{S}_{\textsc{eh}} = \frac{1}{2\kappa} \int \sqrt{-g} \, R \,
\dd^4 x\,,
\end{align}
with $\kappa=8\pi G_\textsc{n}$, describing gravity, and, for the
perfect fluid, the Schutz action~\cite{Schutz:1970my,Schutz:1971ac}
\begin{align}
\mathcal{S}_\text{fluid} = \int \dd^4 x \sqrt{-g} \, 
P \qty(h)\,,
\end{align}
where $h=\qty(\rho+P)/n$ is the enthalpy per particle
($n$ being the number density) which can be recast as
$h^2=-g^{\alpha\beta} \partial_\alpha \phi \partial_\beta \phi$,
thereby defining the velocity potential $\phi$ upon which
the description of the irrotational fluid motion is based
(and which is not to be confused with a scalar matter field).
Setting the explicit form $P=K h^{1+1/w}$ for two arbitrary constants
$K$ and $w$ and assuming the thermodynamical relation $\dd P = h \dd n$ 
yields the barotropic equation of state $P = w\rho$ relating the
fluid pressure $P$ to its energy density $\rho$; the equation of
state parameter $w$ is assumed to be a constant (see 
Ref.~\cite{Malkiewicz_2019} for details).

For a perfect fluid in a FLRW universe, one finds the following
Hamiltonian
\begin{align}
    \Hu_\text{fluid} = c \frac{N}{\pqty{\sqrt{\gamma}}^w}\,
    p_\phi^{1+w} \,,
\label{Hm}
\end{align}
where $\gamma$ is the determinant of the spatial metric, i.e.
$\sqrt{\gamma} = a^3$, $p_\phi = (1 + \frac{1}{w})\sqrt{\gamma}K \mathcal{V}_0 N^{-1/w} \dot{\phi}^{1/w}$
is the momentum associated to the potential $\phi$, $\mathcal{V}_0$ denotes the spatial section volume
$\mathcal{V}_0 = \int\dd[3]{x}$, and the constant
$c \ \in \mathbb{R}$ reads
\begin{align}
c = \frac{w^w }{\qty(K\mathcal{V}_0)^w \qty(1+w)^{1+w}}\,.
\end{align}
In order to obtain a suitable clock $\tau$ for quantization of the
system, we need to bring the total Hamiltonian (gravitation $+$ fluid)
into the deparametrized form in which the gravitational part decouples
from the fluid Hamiltonian, where the latter can then be written as a
linear term $\propto p_\tau \,$.  The matter contribution
\eqref{Hm} naturally provides such a (fluid) clock: setting
the lapse to $N = - a^{3w}$, so that\footnote{We
set the lapse to be always negative to ensure the Hamiltonian
is positive definite. This convenient choice is physically 
irrelevant as it merely accounts for a choice of the time 
direction.} $N/(\sqrt{\gamma})^w=-1$,
it merely remains to perform the canonical transformation
$(\phi, \, p_\phi ) \to (\tau, \, p_\tau)$ 
with 
\begin{align}
 p_\tau = - c p_\phi^{1+w} \quad \hbox{and}
\quad \tau = - \frac{\phi}{c (1+w) p_\phi^w}\,.
\end{align}
Having set the lapse, from now on the time $T$ is given by the fluid
time $\tau$.

The gravitational Hamiltonian for a flat FLRW spacetime reads
\begin{align}
\Hu_\text{grav} = -\frac{\kappa N}{12\mathcal{V}_0 a} p_a^2\,,
\end{align}
where the momentum conjugate to the scale factor is given by $p_a = -
6 \mathcal{V}_0 a \dot{a}/(N\kappa)$, a dot meaning differentiation
with respect to the time $\tau$, i.e.  $\dot{a} \equiv \dd
a/\dd \tau$, which we assume to be finite.
We then find the total Hamiltonian for our lapse choice to read
\begin{align}
    \Hu_\textsc{t} = \Hu_\text{grav} + \Hu_\text{fluid} = \frac{\kappa
    a^{3 w-1}}{12\mathcal{V}_0} p_a^2 + p_\tau\,.
\end{align}
It now suffices to perform another canonical transformation, changing
the variables $(a, p_a)$ into $(q,p)$ (where $w \neq 1$)\footnote{For
the special case $w = 1$ we find that $p
= \sqrt{\kappa/(12\mathcal{V}_0) }\,a \, p_a$ and $q
= \sqrt{12\mathcal{V}_0/ \kappa}\, \log(a)$.} with
\begin{equation}
\begin{split}
q = & \sqrt{\frac{12 \mathcal{V}_0}{\kappa}}
\frac{2 a^{\frac32(1-w)}}{3 (1-w)}\,, \\
p = & \sqrt{\frac{\kappa}{12 \mathcal{V}_0}} a^{\frac12
(3w-1)}p_a \propto a^{\frac32(1+w)} H \,,
\end{split}
\label{qanda}
\end{equation}
where $H\equiv\dot{a} / (Na)$ denotes the Hubble rate.  This leads to
a simplified version of the gravitational Hamiltonian
\begin{equation}
\Hu_\text{grav} =  p^2\,.
\label{Hgrav}
\end{equation}
The gravitational Hamiltonian thus takes the canonical form used
already in Ref.~\cite{Bergeron:2023zzo}.

Solving the classical Hamilton equations of motion for
$\Hu_\text{grav}$, we find that the momentum is constant,
i.e. $p(\tau) =p_0$, so that $\dot{q} =2p_0$, which is easily
integrated as $q= 2 p_0 \qty(\tau - \tau_\textsc{s})$: the classical
universe either grows from or contracts to a singularity at
$\tau=\tau_\textsc{s}$, which is apparent from the relation between
$q$ and the scale factor \eqref{qanda}, assuming that $0<w<1$.

\section{Quantization}
\label{sec:quantisation}

After having defined the classical system above, we turn to its
quantization. Although the reader may find a complete description of
our procedure in Ref.~\cite{Bergeron:2023zzo}, for the sake of
completeness and self-consistency, this section recaps and summarizes
the results obtained there.

In order to quantize the gravitational Hamiltonian obtained above, we
first note that since the scale factor $a$ is defined to be
non-negative, then as long as $w<1$, the phase space variable $q$ is a
non-negative variable, i.e. $q \in \mathbb{R}^+$, whereas
$p \in \mathbb{R}$, see Eq.~\eqref{qanda}: the usual canonical
quantization based on the Weyl-Heisenberg group is thus not
applicable. Instead, we make use of the so-called affine quantization
(see Ref.~\cite{Martin:2021dbz} and references therein), based on the
affine group representing dilations and translations on the real line.

There exists a useful quantization scheme which is based on a basis of
coherent states $\ket{q,p}_{\psi_0}$. They are obtained through a
unitary irreducible representation $\hat{U}(q,p)$ of the two-parameter
affine group acting on the Hilbert space $\mathscr{H}$, namely
\begin{align}
\ket{q,p}_{\psi_0} = \underbrace{\exp \qty( \ii 
\frac{p}{2q} \hat{X}^2)
e^{-\frac12 \ii \ln q \qty( \hat{X} \hat{P} +
\hat{P} \hat{X})}}_{\hat{U}(q,p)} \ket{\psi_0}\,,
\label{eq:cohState}
\end{align}
and they depend on an a priori arbitrary fiducial state
$\ket{\psi_0} \in \mathscr{H}$.

Operators corresponding to phase space functions $f(q,p)$ are then
obtained from the quantization map~\cite{Bergeron:2023zzo}
\begin{align}
\hat{A}_f (\psi_0) = \mathcal{N}_{\psi_0} \int_{\mathbb{R} \times \mathbb{R}^+}
\, \ket{q,p}_{\psi_0}\, f(q,p)\, \null_{\psi_0}\!\!\bra{q,p} \dd q \, \dd p\,,
\label{quantMap}
\end{align}
where $\mathcal{N}_{\psi_0}$ is a normalization factor ensuring
$\hat{A}_1 (\psi_0) = \mathbb{1}$, thereby leading to a first integral
constraint on the fiducial state $\ket{\psi_0}$. Demanding also that
$\hat{A}_q (\psi_0) = \hat{X}$ (with $\hat{X} \psi = x \psi$) and
$\hat{A}_p (\psi_0) = \hat{P}$ (with $p\to \hat{P} \psi =
-\ii\partial_x \psi$), so that $\comm{\hat{A}_q (\psi_0)}{\hat{A}_p
(\psi_0)} = \comm{\hat{X}}{\hat{P}} = \ii$, yields another integral
constraint on $\ket{\psi_0}$ (as shown explicitly in the Appendix).

Upon using the above quantization scheme\footnote{A similar result can
be reached by merely considering the operator ordering ambiguity,
i.e. using that $\forall \alpha \in \mathbb{N}$, $p^2 = x^\alpha p
x^{-2\alpha} p x^\alpha$, classically, which yields the extra term
$\propto \hat{X}^{-2}$ in \eqref{Hquant} upon canonical quantization.}
one ends up with a quantum Hamiltonian $\hat{\Hu}_\text{grav}
= \hat{A}_{p^2} (\psi_0) = \hat{\Hu}_\nu$ depending on one arbitrary
constant $\nu\in \mathbb{R}$, again in principle calculable as an
integral over the fiducial state, as is given in
the Appendix, see Eq.~\eqref{AppP2}. Applying the quantization procedure
as described also in Ref.~\cite{Bergeron:2023zzo}, one finds that
$\hat{\Hu}_\text{grav}$ is given by
\begin{equation}
\hat{\Hu}_\text{grav} = \hat{A}_{p^2} = \hat{\Hu}_\nu \equiv \hat{P}^2 +
\qty(\nu^2 -\frac14) \hat{X}^{-2}\,,
\label{Hquant}
\end{equation}
where $\hat{P} = -\ii \partial_x$ and one demands that $\nu >\frac12$
in order for the potential $\propto \hat{X}^{-2}$ to be repulsive. In
order for the quantum Hamiltonian to be self-adjoint we must have
$\nu^2 - \frac14 \geq \frac34$~\cite{Reed:1975uy,Pouliquen:1999,
Gitman_2010} and thus restrict to $\nu \geq 1$. 
Physically speaking, the value of $\nu$ enters the dynamics of the scale 
factor as discussed in Sec.\,\ref{sec:dBB} but is degenerate with initial conditions
when it comes to physical quantities like the curvature at the bounce or the rate of late time expansion in the  case of a single state universe. 
The quantized
Hamiltonian \eqref{Hquant} is the starting point of the following
analysis; as is detailed in the appendix, this explicit form implies a rescaling of the time parameter $\tau$ by a constant depending on the fiducial state.

In Ref.~\cite{Bergeron:2023zzo}, the authors propose a new class of
coherent states $\ket{q(\tau),p(\tau)} $ for which the initially
arbitrary numbers $q$ and $p$ labeling the basis for the Hilbert space
are turned into time dependent functions that solve the equations of
motion arising from the semi-classical Hamiltonian given by
\begin{equation}
\Hu_\textsc{sc} = \Hu_\textsc{sc}(q,p) = \bra{q,p} \hat{\Hu}_\text{grav}
\ket{q,p} \propto p^2 + \frac{\xi_\nu^2}{q^2}\,,
\label{Hsem}
\end{equation}
where $\xi_\nu$ is a positive-definite function of $\nu$ we fix below
in Eq.~\eqref{xinnu} (see \cite{Bergeron:2023zzo} and the appendix of
Ref.~\cite{Martin:2022ptk} for details).

Solving Hamilton's equations for \eqref{Hsem} in terms of the fluid
time $\tau$, introducing the notation $q_\tau\equiv q(\tau )$ and
$p_\tau\equiv p(\tau )$, one finds
\begin{equation}
\begin{split}
    q_\tau = & q_\textsc{b} \sqrt{1
    + \omega^2 \qty(\tau-\tau_\textsc{b})^2}, \\ p_\tau =
    & \frac12 \dot{q}(\tau)
    = \frac{q_\textsc{b}\, \omega^2 \qty(\tau-\tau_\textsc{b})}{2 \sqrt{1
    + \omega^2 \qty(\tau-\tau_\textsc{b})^2}}\,,
\end{split}
\label{qptau}
\end{equation}
representing a bouncing trajectory with constant energy $E=p_\tau^2
+\xi_\nu^2/q_\tau^2$. The minimum scale factor is determined by the
relation $q_\textsc{b} = \xi_\nu/\sqrt{E}$, while the change of
momentum at the bounce is determined by $\omega = 2 E/\xi_\nu$.
Although the actual bouncing time $\tau_\textsc{b}$ is arbitrary, and
could thus be set to $\tau_\textsc{b}\to 0$ for a single solution, we
shall keep it explicit as we consider more than one such trajectory in
what follows.

Using \eqref{Hquant} and Dirac's correspondence principle
($p_\tau \to \hat{p}_\tau \psi = -\ii\partial_\tau \psi$) transforms
the constraint $\Hu_\textsc{t} = \Hu_\text{fluid} + \Hu_\text{grav}
\simeq 0$ into the time-dependent Schr\"odinger equation
\begin{align}
\ii \partial_\tau \psi (x,\tau) = \underbrace{\qty(-\partial^2_x +
\frac{\nu^2-\frac14}{x^2})}_{\hat{\Hu}_\nu }
\psi (x,\tau)\,,
\label{HnuSch}
\end{align}
whose solution reads $\psi(x,\tau) = e^{-\ii \phi_\tau} \bra{x}
\ket{q_\tau,p_\tau}_n$, with 
\begin{align}
e^{-\ii \phi(\tau)} = \qty( \frac{\xi_\nu -
\ii p_\tau q_\tau}{\xi_\nu 
+ \ii p_\tau q_\tau})^{\beta_n/(4\xi_\nu)}
\label{phasePhi}
\end{align}
and
\begin{align}
\ket{q_\tau,p_\tau}_n = 
\exp \qty( \ii \frac{p_\tau}{2q_\tau} \hat{x}^2)
e^{-\frac12 \ii \ln q_\tau \qty( \hat{x} \hat{p} +
\hat{p} \hat{x})}\ket{\Phi_n}\,,
\end{align}
in which the new fiducial state $\ket{\Phi_n}$, $n \in \mathbb{N}$, is
taken to be an eigenvector of the auxiliary Hamiltonian
$\hat{\Hu}_\text{aux} = \hat{\Hu}_\nu + \xi_\nu^2 \hat{X}^2$ with
eigenvalue $\beta_n$.  Note that the numerical value of $\xi_\nu$ in
$\hat{\Hu}_\text{aux}$ is the same as that in the semi-classical
$\Hu_\textsc{sc}$ in \eqref{Hsem}. For details on the introduction of
$\hat{\Hu}_\text{aux}$, see \cite{Bergeron:2023zzo}.  In the
$x-$representation, the unitary operator in the above relation yields
\begin{align}
\braket{x}{q_\tau,p_\tau}_n = 
\frac{1}{\sqrt{q_\tau}} \exp \qty( \ii \frac{p_\tau}{2q_\tau} x^2)
\Phi_n \qty(\frac{x}{q_\tau}).
\end{align}
The spectrum of $\hat{\Hu}_\text{aux}$ is known, its eigenvalues
$\beta_n = 2\xi_\nu \qty(2n+\nu+1)$ are associated with the normalized
eigenvectors $\Phi_n(x) = \braket{x}{\Phi_n}$ reading
\begin{equation}
\Phi_n(x) = \sqrt{\frac{2n!}{\Gamma\qty(n+\nu+1)}} \xi_\nu^{\frac{\nu+1}{2}}
x^{\nu+\frac12} L^{(\nu)}_n \qty( \xi_\nu x^2) e^{-\frac12 \xi_\nu x^2},
\label{Phin}
\end{equation}
where $L_n^{(\nu)}$ is an associated Laguerre
polynomial~\cite{olver10}.

The solution is then fully determined by the last requirement that its
expectation value follows the semi-classical trajectory, i.e. by
demanding that
$\null{}_n\!\bra{q_\tau,p_\tau}\hat{X}\ket{q_\tau,p_\tau}_n = q_\tau$,
which imposes the choice
\begin{equation}
\xi_\nu \to \xi_{\nu,n} = \qty{ \frac{n!}{\Gamma\qty(n+\nu+1)}
\int_0^\infty y^{\nu+\frac12}
\qty[ L_n^{(\nu)}(y) ]^2 e^{-y} \dd y }^2\,.
\label{xinnu}
\end{equation}
It turns out that the same requirement on the momentum, i.e.
$\null{}_n\!\bra{q_\tau,p_\tau}\hat{P}\ket{q_\tau,p_\tau}_n = p_\tau$,
yields the exact same constraint on the value of $\xi_\nu$. 
The integral in \eqref{xinnu} can be written as
\begin{equation}
\xi_{\nu,n} = \frac{\Gamma^2 \qty(\nu +\frac{3}{2})}
{\Gamma^2 \qty(\nu + n + 1)} \qty( \nu^{2n} + \cdots)\,,
\label{xinu}
\end{equation}
where the unwritten terms form a complicated polynomial of order
$2n-1$ with $2n-1$ terms whose coefficients cannot be
expressed in simple closed form in terms of $n$ and $\nu$; for the case $n=0$ we focus on below
Eq.~\eqref{xinu} reduces to $\xi_\nu = \Gamma^2\qty(\nu+
\frac32)/ \Gamma^2\qty(\nu+1)$.

Our basis of states therefore consists of the set $\qty{
\ket{q_\tau,p_\tau}_n}$ labeled by the energy level $n$ at
which the fiducial state is chosen; they also depend on the
amplitude of the quantum barrier $\nu$.

\section{Physical background state}
\label{sec:bckgS}

The expectation value of each of the coherent states
$\ket{q_\tau,p_\tau}_n$ defined above follows a definite
semi-classical trajectory $q_\tau$ as given by Eq.~\eqref{qptau},
which depends on $q_\textsc{b}$ and $\omega$.  These two quantities
are determined by the conserved energy $E$ and the constant $\xi_\nu$,
the latter of which is determined by the Hamiltonian; hence, both of
them are fixed once $E$ is given.  The time at which the bounce takes
place $\tau_\textsc{b}$ is fully arbitrary, being often, as we
discussed above, set to zero to fix the origin of time.  We therefore
write a generic coherent state as $\ket{q_\tau,p_\tau}_n \equiv
\ket{E,\tau_\textsc{b}}_n$.

Since these states provide a natural trajectory \eqref{qptau} for the
scale factor from $\langle \hat{q}\rangle$, it seems adequate to pick
any one of them to define the time development of the background in
the early Universe. The usual approach~\cite{Kiefer:2025udf} consists
in finding a background state, expected to be as classical
(i.e. peaked) as possible, leading to a semi-classical trajectory over
which one subsequently evolves perturbations, which are often assumed
to be initially in a quantum vacuum state. In other words, one
considers the Universe ($\ket{\Psi_\textsc{u}}$) evolution to be
described by background ($\ket{\Psi_\textsc{b}}$) and perturbation
($\ket{\Psi_\textsc{p}}$) states such that the Born-Oppenheimer
approximation $\ket{\Psi_\textsc{u}} = \ket{\Psi_\textsc{b}}
\otimes \ket{\Psi_\textsc{p}}$ should hold at all times.

Thus, the usual treatment of the early Universe contains two implicit
assumptions, namely the Born-Oppenheimer approximation as well as
postulating a close-to-classical background state. The former
hypothesis was put to the test in
Refs.~\cite{Bergeron:2024art,Bergeron:2025eda} leading to the
conclusion that entanglement between the various perturbation states
of different backgrounds could induce non-gaussianities even in an
otherwise purely gaussian theory.

We now wish to examine the second assumption of a simple almost
classical state.  For this, we construct the most general background
state from the basis of coherent states which individually represent a
would-be semi-classical trajectory as the superposition
\begin{equation}
\ket{\Psi_\textsc{b}} = \mathcal{N}(\tau)
\sum_{n\in\mathbb{N}} \int \dd E \int \dd
\tau_\textsc{b} W_n\qty(E,\tau_\textsc{b}) \ket{E,\tau_\textsc{b}}_n\,,
\label{GenBck}
\end{equation}
where the weight function $W_n\qty(E,\tau_\textsc{b})$ is arbitrary
and $\mathcal{N}(\tau)$ a time-dependent normalization.  Since each
state $\ket{E,\tau_\textsc{b}}_n$ independently solves the
Schr\"odinger equation \eqref{HnuSch}, so does the state
\eqref{GenBck}.

Since a priori one has no guiding principle for the form of the weight
function $W_n\qty(E,\tau_\textsc{b})$, one can only propose various
forms and investigate their consequences to obtain indications for its
reconstruction.  Such an ambitious program is out of reach, so we
focus on the treatment of the simplest potentially interesting cases,
naturally excluding the single state case $W_n\qty(E,\tau_\textsc{b})
= \delta_{n,n_0} \delta\qty(E-E_0) \delta\qty(\tau_\textsc{b}-\tau_0)$,
which has already been explored in detail
elsewhere~\cite{Peter:2002cn,Peter:2006hx,Martin:2021dbz}.  As the
simplest extension, we consider sum of $N$ semi-classical basis
states, each entering with a potentially different weight
$\frak{w}_a\in\mathbb{C}$, i.e. we consider a weight function such
that
\begin{widetext}
\begin{align}
\mathcal{N}(\tau) W_n\qty(E,\tau_\textsc{b}) \to \mathcal{N}_N (\tau)
\sum_{a=1}^{N} \frak{w}_a \delta_{n,n_a} 
\delta\qty(E-E_a) \delta\qty(\tau_\textsc{b}-\tau_{\textsc{b},a})\,,
\end{align}
the normalization coefficient $\mathcal{N}_N\in\mathbb{R}$ being given
by
\begin{align}
\mathcal{N}_N (\tau)= \qty( \sum_{a,b=1}^{N}\, \frak{w}_a^\star
\frak{w}_b \, \null_{n_a}\!\!\braket{E_a,\tau_{\textsc{b},a}}
{E_b,\tau_{\textsc{b},b}}_{n_b})^{-1/2} \,,
\end{align}
ensuring that $\braket{\Psi_\textsc{b}}{\Psi_\textsc{b}}=1$.  Note
that $\mathcal{N}_N$ actually depends on time because even though each
basis state is normalized, i.e.
$\null_{n_a}\!\!\braket{E_a,\tau_{\textsc{b},a}}{E_a,\tau_{\textsc{b},a}}_{n_a}
= 1$, the basis of coherent states we are using is however
overcomplete so that
$\null_{n_a}\!\!\braket{E_a,\tau_{\textsc{b},a}}{E_b,\tau_{\textsc{b},b}}_{n_b} \not=
0$.  From now on, for notational simplicity, we name a given state by
a single label $a
\equiv \qty{n_a,E_a,\tau_{\textsc{b},a}}$, namely
$\ket{E_a,\tau_{\textsc{b},a}}_{n_a}
\to \ket{a}$ (this labeling will only occur in this section, so no
confusion with the scale factor should arise).

In the $x-$representation, the wave function $\psi_a (x) =
\braket{x}{a} = \braket{x}{E_a,\tau_{\textsc{b},a}}_{n_a} $ corresponding to
one of our states reads~\cite{Bergeron:2023zzo}
\begin{equation}
\psi_a (x) = \sqrt{\frac{2 \,\, n_a!}
{\Gamma(\nu+n_a+1)}} \qty( \frac{\xi_{\nu, n_a} - \ii q_a p_a}
{\xi_{\nu, n_a} + \ii q_a p_a} )^{\frac12 \qty(2n_a+\nu+1)}
\xi_{\nu,n_a}^{\frac{\nu+1}{2}} \, \frac{x^{\nu+1/2}}{q_a^{\nu+1}} 
L_{n_a}^\nu\qty( \xi_{\nu,n_a} \frac{x^2}{q_a^2} )
\exp \qty[ - \frac12 \qty( \xi_{\nu,n_a} - \ii \, q_a p_a)
\frac{x^2}{q_a^2} ]\,,
\label{GenStateX}
\end{equation}
with $L_{n_a}^\nu$ a Laguerre polynomial and $\qty{q_a,p_a}$ the
solutions to the semi-classical trajectories as given in
Eqs.~\eqref{qptau} with parameters $q_\textsc{b} \to q_{\textsc{b},a}
= \xi_{\nu,n_a} /\sqrt{E_a}$, $\omega \to \omega_a =
2E_a/\xi_{\nu,n_a}$ and bouncing time
$\tau_\textsc{b} \to \tau_{\textsc{b},a}$.

Eq.~\eqref{GenStateX} permits to calculate the scalar product
\begin{align}
\braket{a}{b} = \null_{n_a}\!\!\braket{E_a,\tau_{\textsc{b},a}}
{E_b,\tau_{\textsc{b},b}}_{n_b} =
\int_0^\infty \null_{n_a}\!\!\braket{E_a,\tau_{\textsc{b},a}}{x}
\braket{x}{E_b,\tau_{\textsc{b},b}}_{n_b}\dd x = \int_0^\infty
\psi_a^\star (x) \psi_b(x)\dd x\,,
\end{align}
and we find
\begin{equation}
\braket{a}{b} = 2 \sqrt{\frac{n_a!}{\Gamma(\nu+n_a+1)}\frac{n_b!}
{\Gamma(\nu+n_b+1)}} \qty( \frac{\xi_{\nu, n_a} + \ii q_a p_a}
{\xi_{\nu, n_a} - \ii q_a p_a} )^{\frac12 \qty(2n_a+\nu+1)}
\qty( \frac{\xi_{\nu, n_b} - \ii q_b p_b}
{\xi_{\nu, n_b} + \ii q_b p_b} )^{\frac12 \qty(2n_b+\nu+1)}
\qty( \frac{\sqrt{\xi_{\nu,n_a}\xi_{\nu,n_b}}}
{q_a q_b})^{\nu+1} I_{ab}\,,
\label{EvalScalarAB}
\end{equation}
with the integral $I_{ab}$ given by
\begin{align}
I_{ab} \equiv \int_0^\infty 
L_{n_a}^\nu\qty( \xi_{\nu,n_a} \frac{x^2}{q_a^2} )
L_{n_b}^\nu\qty( \xi_{\nu,n_b} \frac{x^2}{q_b^2} )
e^{-z_{ab} x^2} x^{2\nu +1}\,\dd x
\label{Iab}
\end{align}
\end{widetext}
and
\begin{align}
z_{ab} \equiv \frac12 \qty[ \frac{\xi_{\nu,n_a}}{q_a^2}
+ \frac{\xi_{\nu,n_b}}{q_b^2} + \ii \qty( \frac{p_a}{q_a}
-\frac{p_b}{q_b})]\,.
\end{align}
Expanding the Laguerre polynomials explicitly and performing the
integrals, one gets the final form of the scalar product through
\begin{align}
\begin{split}
I_{ab} = & \frac12 \sum_{\ell=0}^{n_a} \sum_{m=0}^{n_b}
\frac{\qty( \nu+\ell+1)_{n_a-\ell}}{\qty(n_a-\ell)! \, \ell !}
\frac{\qty( \nu+m+1)_{n_b-m}}{\qty(n_b-m)! \, m!}\\
& \times \qty(\frac{\xi_{\nu,n_a}}{q_a^2})^\ell
\qty(\frac{\xi_{\nu,n_b}}{q_b^2})^m
\frac{\Gamma\qty(\nu + \ell+ m +1)}{z_{ab}^{\nu + \ell+ m +1}}\,,
\end{split}
\label{Iabeval}
\end{align}
where $(\alpha)_n$ are Pochhammer symbols~\cite{olver10} given by
$(\alpha)_n = \alpha (\alpha +1)\cdots (\alpha + n -1)$ and
$(\alpha)_0 = 1$.

In what follows, we shall restrict attention to the simplest possible
case for which the wave functions are produced from the ground state
of the auxiliary Hamiltonian and thus set $n_a=n_b=0$; this assumption
also stems from the fact that with $n\neq 0$, the corresponding
coherent state shows $n+1$ peaks so the expectation value does not
correspond to such a peak, contrary to the usual understanding that
the background state is ``mostly classical'', meaning its probability
distribution is peaked around its expectation value with comparatively
small variance. Note that since we investigate precisely the relevance
of the above hypothesis, there is a priori no reason to restrict
attention to states with $n=0$.  The effect we wish to emphasize is
however already present in $n=0$ superposition states, which are also
the ones employed in Refs.~\cite{Bergeron:2024art, Bergeron:2025eda}.

Setting $n =0$, the associated Laguerre polynomials appearing
in \eqref{GenStateX} are simply given by unity, i.e. $L_0^\nu = 1$.
The scalar product between two basis states then simplifies to
\begin{align}
\begin{split}
\braket{a}{b} = & \ 2^{1+\nu} \qty[ \qty( \frac{q_a}{q_b}
+ \frac{q_b}{q_a})^2 + \frac{\ii}{\xi_\nu} \qty( p_a q_b - p_b
q_a)]^{-(1+\nu)}\\ &\times \qty( \frac{\xi_\nu+\ii q_a
p_a}{\xi_\nu-\ii q_a p_a})^{\frac12\qty(1+\nu)}
\qty( \frac{\xi_\nu-\ii q_b p_b}{\xi_\nu+\ii q_b p_b})^{\frac12\qty(1+\nu)}\,,
\label{abt}
\end{split}
\end{align}
in which, by virtue of \eqref{xinu}, $\xi_\nu \equiv
\xi_{\nu,0}$ takes
the value $\xi_\nu = \Gamma^2\qty(\nu+\frac32)/
\Gamma^2\qty(\nu+1)$. 
Eq.~\eqref{abt} makes explicit that the basis is overcomplete as the
scalar product between different basis states $b\neq a$ is
non-vanishing. In fact, this overlap in time-independent and can be
calculated explicitly as
\begin{align}
\braket{a}{b} = \qty( \frac{2\sqrt{r_{ab}}}
{1+r_{ab} + \ii \, r_{ab} \omega_a\Delta\tau} )^{1+\nu}\,,
\label{ab0}
\end{align}
in which we set $r_{ab}\equiv E_a/E_b$, $\Delta\tau
= \tau_{\textsc{b},a}- \tau_{\textsc{b},b}$ and $\omega_a = 2E_a
/\xi_\nu$.  Note that since one demands that $\nu>1$ in order to
ensure the Hamiltonian \eqref{Hquant} be self-adjoint,
$|\braket{a}{b}|$ increases with $r_{ab}$ and decreases with
$\Delta \tau$ and $\omega_a$. For the same set of bounce times and
$\omega_a$, the overlap between states is thus higher if they have
similar energies.

In our numerical examples, we set $\nu$ to its minimal physically
permitted value $\nu = 1$, so that $\xi_\nu = 9\pi/16$ and the real
part of the integral
\eqref{ab0} simplifies to
\begin{align}
\Rez \braket{a}{b} = 4r_{ab} \, 
\frac{\qty(1+r_{ab})^2 - r_{ab}^2 \omega_a^2 \Delta\tau^2}
{\qty[\qty(1+r_{ab})^2 + r_{ab}^2 \omega_a^2 \Delta\tau^2]^2}\,,
\label{Reab}
\end{align}
while the imaginary part becomes
\begin{align}
\Imz \braket{a}{b} = - 
\frac{8 r_{ab}^2 \qty(1+r_{ab}) \omega_a \Delta\tau}
{\qty[\qty(1+r_{ab})^2 + r_{ab}^2 \omega_a^2 \Delta\tau^2]^2}\,,
\label{Imab}
\end{align}
both of which are required to normalize the full wave function.

As a concrete example to be investigated in detail in
Sec.~\ref{sec:num}, we restrict our attention to the two wave function
case, i.e. a biverse similar to~\cite{Bergeron:2024art}: the
normalized background state we consider thus reads
\begin{align}
\ket{\Psi_\textsc{b}} = \mathcal{N}_2 \, \qty( \ket{0} +
\rho e^{-\ii\delta} \ket{1}),
\label{PhiB2}
\end{align}
where we discard the irrelevant global phase and set $\frak{w}_0 =1$
and $\frak{w}_1 = \rho e^{-\ii\delta}$, with $\rho,\delta
\in\mathbb{R}$. With the explicit parametrization of \eqref{PhiB2},
the normalization $\braket{\Psi_\textsc{b}} =1$ implies
\begin{align}
\mathcal{N}_2 = \qty[ 1+\rho^2 +2\rho \qty( \cos\delta \, \Rez\braket{0}{1}
+\sin\delta \, \Imz\braket{0}{1})]^{-1/2}\,.
\end{align}
Overall, starting from a simple state with definite energy $E_0$ and
bounce time $\tau_{\textsc{b},0}$, which can be chosen as
$\tau_{\textsc{b},0}=0$ without loss of generality, the simplest
possible extension for the background is thus described by four
parameters, namely the relative contribution in the wave function
(amplitude $\rho$ and phase $\delta$), the ratio between the energies
involved $r \equiv r_{10} = E_1/E_0$ along the coherent state
trajectory, and finally the bounce time difference
$\tau_{\textsc{b},1}-\tau_{\textsc{b},0}=\Delta\tau$.

\section{Scale factor time development}
\label{sec:dBB}

Ultimately, any physical consequences of the superposition states of
the form \eqref{GenBck} or \eqref{PhiB2} introduced in the previous
section would manifest in cosmological perturbations.  As discussed in
the introduction, the evolution of these perturbations is usually
calculated over a background obtained by assuming a Born-Oppenheimer
approximation. Within this approximation, one defines a trajectory as
the expectation value of the scale factor operator in a background
state, which is thus required to be sufficiently peaked. The resulting
trajectory is subsequently plugged into the perturbation mode
evolution equation. For a single coherent state of the form given
in \eqref{GenStateX}, or any other case where the peak of $|\Psi|^2$
clearly follows a trajectory in phase space, this might appear to be
an unambiguous procedure, but for the composite states discussed in
the previous section this is less clear.

One could take the viewpoint that the wave function needs to be
``projected'' onto one of the states appearing in \eqref{GenBck} and
that each background state sources its respective perturbations. This
viewpoint was taken in e.g. ~\cite{Bergeron:2024art,Bergeron:2025eda},
where the behavior of perturbations is influenced by the existence of
several background states due to a relaxation of the Born-Oppenheimer
approximation, which is explicitly not the case for our considerations
below.

Instead, here we make use of quantum trajectories as proposed already
in \oldstylenums{1927} by de Broglie~\cite{deBroglie:1927} and
subsequently developed by Bohm~\cite{Bohm:1951xw,Bohm:1951xx}, and
thus consider the eikonal approach, which gives an unambiguous
trajectory for the entire background evolution for any choice of
quantum state that satisfies the Schr\"odinger equation (see
Refs.~\cite{Pinto-Neto:2013toa,Pinto-Neto:2018zvn} for its application
in our framework of quantum cosmology). One can then study the
evolution of perturbations on the background determined by this single
trajectory using the Born-Oppenheimer approximation. In the following,
we introduce the quantum trajectories and compute explicitly the
solutions for a quantum state described by a single wave function as
given in \eqref{GenStateX}; we return to the biverse case in the next
section and consider perturbations in Sec.\,\ref{sec:perturbations}.

The trajectories $x(\tau)$ postulated by the eikonal approach are
driven by the wave function $\Psi_\textsc{b}$.  The imaginary part of
the Schr\"odinger equation in the $x-$representation takes the form of
a conservation equation for the probability density
$\abs{\Psi_\textsc{b}(x,\tau)}^2$, namely
\begin{align}
\frac{\partial}{\partial\tau} \abs{\Psi_\textsc{b}(x,\tau)}^2
+ \frac{\partial}{\partial x} \qty[ 2 \abs{\Psi_\textsc{b}(x,\tau)}^2
\frac{\partial S(x,\tau)}{\partial x}] =0\,,
\label{conservation}
\end{align}
with $S(x,\tau)$ the phase of the wave function. Assuming
$\abs{\Psi_\textsc{b}(x,\tau)}^2$ represents a statistical
distribution, one can naturally interpret the gradient of the phase as
a ``velocity'' in the above conservation equation, to obtain an
equation of motion for the phase space trajectories
 
\begin{align}
\frac{\dd x}{\dd\tau} = 2 \partial_x S =
\frac{\Psi_\textsc{b}^\star \partial_x
\Psi_\textsc{b} - \Psi_\textsc{b} \partial_x \Psi_\textsc{b}^\star}{\ii
\abs{\Psi_\textsc{b}}^2}
= -\ii \partial_x \ln \frac{\Psi_\textsc{b}}{\Psi_\textsc{b}^\star}\,,
\label{dotxdBB}
\end{align}
where we used the substitution $x\to x(\tau)$.  To see that $x(\tau)$
indeed provides a trajectory through phase space, consider that
$\hat{q} \Psi(x) = x \Psi(x)$ for $\hat{q}$ obtained
from \eqref{quantMap}.  The transformation~\eqref{qanda} shows that
the $x(\tau)$ trajectory is related to the scale factor through
\begin{align}
x(\tau) = \sqrt{\frac{12 \mathcal{V}_0}{\kappa}} 
\frac{2 \, a^{\frac{3}{2}(1-w)}}{3(1-w)} \equiv \lambda
a^{\frac{3}{2}(1-w)}\,,
\label{xa}
\end{align}
thereby leading to a scale factor trajectory $a(\tau)$.  Note that,
depending on the state $\Psi_\textsc{b}$, the evolution of the
trajectories $x(\tau)$ can strongly differ from the semi-classical
dynamics \eqref{qptau}; in particular, for the biverse case the
trajectory dynamics will be very different to the semi-classical
solutions.

From \eqref{dotxdBB} and the Schr\"odinger equation it follows that
\begin{align}
\ddot{x} = - 2 \pd_x \qty[V(x) + Q(x,t)]\,,
\end{align}
for a general Hamiltonian of the form $\hat{H}= -\pd_x^2 + V(x)$ and
with the so-called quantum potential $Q(x,\tau)$ defined by
\begin{align}
Q(x,\tau) \equiv -\frac{1}{|\Psi_\textsc{b}|}
\pdv[2]{|\Psi_\textsc{b}|}{x}\,.
\label{Qpot}
\end{align} 
The quantum trajectories follow the dynamics generated by the
classical Hamiltonian $H_\text{cl} = p^2 + V(q)$ when $Q \to 0 $.

In the single wave function case, the solution \eqref{GenStateX}
yields the following expression for the phase
\begin{align}
S_a = -\phi_a + \frac{p_a x^2}{2 q_a} = -\phi_a + \frac{\omega_a^2
x^2 \qty(\tau-
\tau_{\textsc{b},a})}{2 \qty[ 1 + \omega_a^2 \qty(
\tau - \tau_{\textsc{b},a})^2]}\,,
\label{Ssingle}
\end{align}
with $\phi_a$ given in \eqref{phasePhi} implying 
$\dot{x}/
x = \dot{q}/q $
using \eqref{qptau}. 
These trajectories are solved by 
\begin{align}
x(\tau) = x(0) \sqrt{1 + \omega_a^2 
\qty(\tau - \tau_{\textsc{b},a})^2} \equiv x_0 \frac{q_a(\tau)}{q_\textsc{b}}\,.
\label{xdBBSingle}
\end{align}
and follow the bouncing semi-classical solution for $q(\tau)$ given
in \eqref{qptau}. The value of $x(0)=x_0$, i.e. the value of the
trajectory at its minimum, is a free parameter that leads to different
trajectories, reproducing exactly the solution \eqref{qptau} for
$x_0\to q_\textsc{b}$. For large values of $|\tau|$, i.e. away from
the bounce, we find $\dot{x} = \pm x_0 \omega_a $, such that the
contraction/expansion rate of the Universe is determined by the
initial conditions and differs for each trajectory.

The quantum potential originating from \eqref{GenStateX} for $n=0$
reads
\begin{align}
    Q(x, \tau) = \frac{2 \xi_\nu (\nu +1) }{q^2} - \frac{\xi^2_\nu x^2
    }{q^4}- \frac{\nu^2 - \frac14 }{ x^2}\,,
\end{align}
such that we find, for $\hat{H}_\nu$ given in \eqref{HnuSch},
\begin{align}
\ddot{x} = - 2 \frac{\partial}{\partial x} \qty[
\frac{\nu^2-\frac14}{x^2} + Q(x,\tau)] = \frac{ 4 \xi_\nu^2}{q(\tau)^4}
x.
\label{fmaq}
\end{align}
Solving the second order equation for the trajectories leads to two
solutions, only one of which satisfies the first order equation and is
proportional to $q(\tau)$. Similar to the case of GR, where the first
Friedmann equation and the continuity equation suffice to solve the
dynamics, the above equation is not necessary to obtain
trajectories. However, it is useful to distinguish between the quantum
and (semi-)classical regime, as the trajectories satisfy the dynamics
generated by the (semi-)classical Hamiltonian, when the quantum
potential $Q$ vanishes.

The guiding equation \eqref{dotxdBB}, once rewritten in terms of the
scale factor $a(t)$ depending on cosmic time $t$, can be understood as
the quantum-corrected version of the first Friedmann equation.
Recalling that the Hubble rate is defined as $H = a^{-1} \dd a/\dd t
= \dot{a}/(N a)$ and using \eqref{xa} as well as $N\dd\tau = \dd t$,
we obtain $H = \frac{2 }{3 (1-w)}\frac{\dot{x}}{N x}\,,$ such that the
first Friedmann equation can be written as
\begin{align}
    H^2 =& \frac{\kappa}{3} \rho = \frac{ 4}{9
    (1-w)^2} \frac{\dot{x}^2}{N^2 x^2} \propto \frac{(\pd_x
    S)^2}{a^{3(1+w)}} \propto \frac{(\pd_a
    S)^2}{a^4} \,.  \label{Hxdot}
\end{align} 
One then recovers the usual Friedmann equation for a universe
dominated by a fluid with equation of state $w$ provided the phase
gradient $\pd_x S$ is asymptotically constant, or equivalently, the
phase behaves asymptotically as $S(a) \sim a^{\frac32 \qty(1-w)}$.

In a fashion similar to \eqref{dotxdBB}, Eq.~\eqref{fmaq} can be
interpreted as the second Friedmann equation, which in classical
cosmology reads $\dot{H} = - \frac12 \kappa N (\rho + P) =
-\frac{3}{2}(1+w) N H^2 \,.$ From \eqref{Hxdot} we have
\begin{align}
    \dot{H} = \frac{2 }{3 (1-w)} \frac{\ddot{x}}{N x}
    - \frac{3}{2}(1+w) N H^2\,,
\end{align}
where we used $N\propto a^{3w}$.  Quantum corrections therefore become
irrelevant whenever when the first term $\propto \ddot{x} \to 0$,
i.e. in our case, whenever $|x(\tau)|$ large and the quantum potential
vanishes, as is indeed the case for the classical solution and for the
single state case far away from the bounce.

The Ricci scalar $R(\tau)$ of a flat FLRW spacetime in terms of the
trajectory $x(\tau)$ reads
\begin{align}
 R(\tau) = 4 \qty(\frac{x}{\lambda
 })^{-\frac{4}{1-w}} \qty[ \frac{\ddot{x}}{(1-w)x} + \frac{ (1- 3
 w) \dot{x}^2}{3 (w-1)^2 x^2}]\,.
\end{align}
At a minimum, i.e.  a bounce, one has $\dot{a} = 0$, so that $\dot{x}
= 0$.  For a single trajectory of the form \eqref{xdBBSingle} we have
$\ddot{x} = x_0 \omega^2 >0$ at the bounce and thus find a maximal
curvature scale $R_\textsc{b}$ given by
\begin{align}
    R_\textsc{b} = \frac{4 \omega ^2}{1-w} \left(\frac{x_0}{\lambda
   }\right)^{-\frac{4}{1-w}} \propto \omega^2(1-w)^{-\frac{3w+1}{(1-w)}}
   x_0^{-\frac{4w}{1-w}}\,,
\end{align}
such that the curvature at the bounce is determined by the initial
conditions and the equation of state parameter.  Assuming $0<w<1$, we
find the Ricci scalar at the bounce increases for small values of
$x_0$, large values of $\omega$ (which translates into large values of
$E$) and diverges as $w\to 1$ (we recall that $w=1$ is a special case
that needs to be treated separately).

In the standard treatment that assumes the Universe to be in a single
coherent state of the form \eqref{GenStateX}, the effective scale
factor describing the evolution of a quantum corrected universe can be
assumed to follow the peak of the Gaussian wave packet (corresponding
to the expectation value), which for the states we used here traces
exactly the semi-classical solutions given by
Eq.~\eqref{qptau}~\cite{Bergeron:2023zzo}.  As we saw, the
trajectories \eqref{xdBBSingle} happen to be merely proportional to
those stemming from the coherent state.  Since the scale factor
normalization is irrelevant in the vanishing spatial curvature case
under consideration here, we can safely conclude that predictions for
the perturbation spectra obtained using either semi-classical or full
quantum trajectories will be exactly the same.  As discussed in the
remainder of the paper, this situation drastically changes when
considering trajectories of superposition states.

\section{Numerics}
\label{sec:num}

We now consider the trajectories for simple superposition states of
the form \eqref{PhiB2}.  In this case, the trajectory
equation \eqref{dotxdBB} has no known general analytic solution,
unlike for the trivial case of a single state discussed above, and
thus requires numerical evaluation.  For this purpose, we introduce
dimensionless variables
\begin{align}
\tilde{x} = \sqrt{E_0} x\,, \quad \tilde{q} = \sqrt{E_0} q\,,
\quad \hbox{and} \quad \tilde{\tau} = E_0\tau\,,
\label{tildes}
\end{align}
defined with respect to the reference energy $E_0$ appearing in the
state $\ket{\Psi_\textsc{b}}$ in \eqref{PhiB2}. The resulting wave
function is also rescaled as $\tilde{\Psi}_\textsc{b} \qty(\tilde{x})
= E_0^{-1/4} \Psi_\textsc{b} \qty(x)$, ensuring its normalization
$\int |\tilde{\Psi}_\textsc{b} \qty(\tilde{x})|^2 \dd{\tilde{x}} =\int
|\Psi_\textsc{b} (x)|^2 \dd{x} =1$.  The trajectories are then
determined by the initial condition
$\tilde{x}(\tilde{\tau}_\text{i})$ and the following four
parameters: the ratio between the energies $r$ and the difference in
bounce times $\Delta \tilde{\tau}$, as well as the relative amplitude
$\rho$ and phase $\delta$ of the two wave functions appearing in the
biverse state [see also below \eqref{ab0} and \eqref{PhiB2}]. In
practice, and since there is no risk of confusion, we omit the tildes
in the discussion of our numerical calculations below.

We solve the equation of motion for the trajectories~\eqref{dotxdBB}
in terms of the rescaled dimensionless variables \eqref{tildes} above,
thus obtaining numerical trajectories $\tilde{x} \qty(\tilde{\tau})$.
To ensure the accuracy of the numerical solutions, we verify that the
Schr\"odinger equation is indeed satisfied along each of these
trajectories. In practice, we evaluate the imaginary and real parts of
the Schr\"odinger equation along a trajectory, which are respectively
given by the conservation equation
\eqref{conservation},
as well as the Hamilton-Jacobi
equation
\begin{align}
\pdv{S}{\tau} + \qty( \pdv{S}{x})^2 + Q + \frac{\nu^2-\frac14}{x^2} =0\,,
\label{HJ}
\end{align}
with $Q$ given by \eqref{Qpot}.  We evaluate the functions $S(x,\tau)$
and $|\Psi_\textsc{b}(x,\tau)|$, by replacing relevant quantities
appearing by their tilde counterparts as well as $\tilde{x}
\to \tilde{x} (\tilde{\tau})$; this allows us
to provide an estimate of the numerical accuracy. For all results
shown below, both the conservation \eqref{conservation} and
the Hamilton-Jacobi \eqref{HJ} equations are satisfied with
an error of at most $10^{-15}$.

\begin{figure}[h]
\includegraphics[scale=0.5]{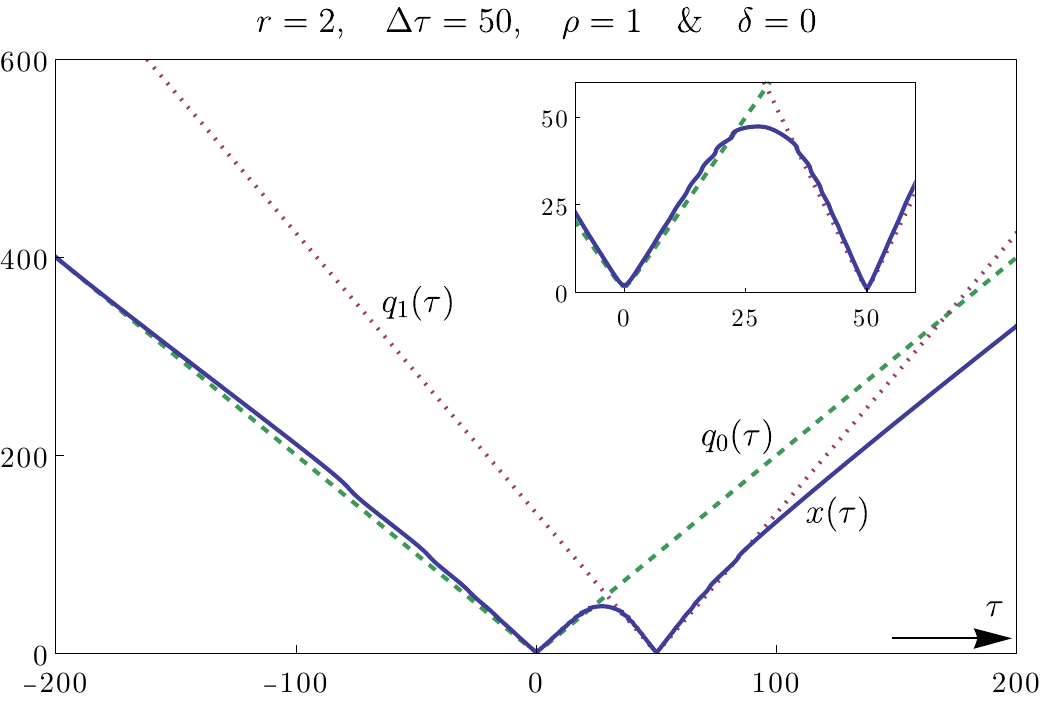}
\caption{Time development of a trajectory for the 
    two-component state \eqref{PhiB2} with parameters $r=2$,
    $\Delta\tau=50$, same amplitude contribution $\rho=1$, and no
    phase $\delta=0$. The reference semi-classical trajectory
    $q_0(\tau)$ appears as the dashed line, while $q_1(\tau)$ is the
    dotted line; the full line shows the actual trajectory $x (\tau)$,
    chosen with initial condition $x (\tau_\text{i}) =
    q_0(\tau_\text{i})$.  We set initial conditions at
    $\tau_\text{i} = - 300$, where the trajectory evolution is close
    to the single state case, but already feels some effects of the
    other wave function.  The insert shows a zoom on the time interval
    separating the two semi-classical bounces. Although not obvious 
    on the figure because of the scale chosen to emphasize the long time
    behavior, all trajectories are regular with a minimum value
    given by $q_0(\tau_{\textsc{b},0}) = q_1(\tau_{\textsc{b},1}) =  \xi_1 = 9\pi/16 \simeq 1.77$. The trajectory $x
    (\tau)$ follows the bounces of the two semiclassical trajectories and exhibits the typical oscillatory behavior characteristic
    of quantum trajectories and emphasized in
    Fig.~\ref{p_r2tau50rho1delta0}.}
\label{x_r2tau50rho1delta0}
\end{figure}

\begin{figure}[t]
\includegraphics[scale=0.5]{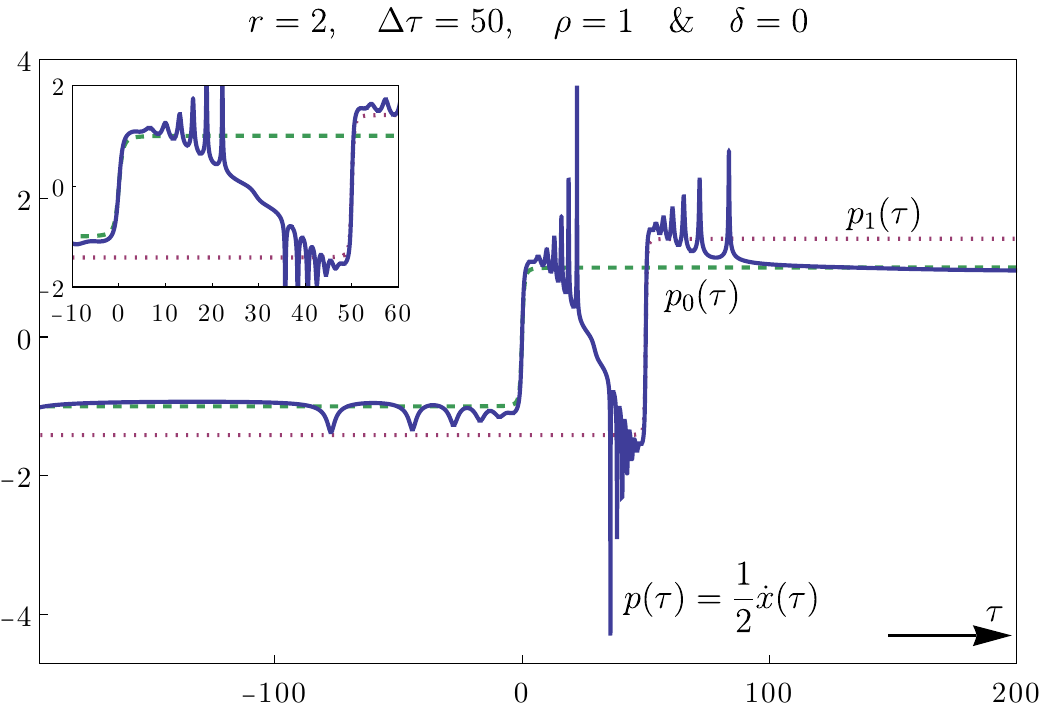}
\caption{Time development of the momentum for the 
    two-component state \eqref{PhiB2} with the same parameters as in
    Fig.~\ref{x_r2tau50rho1delta0} and the same conventions. The
    oscillations along the trajectory translate into large but finite
    peaks in the momentum (they appear sharp in the figure for reasons
    pertaining to image resolution). The insert zooms on the central
    region showing again oscillating but smooth transitions.}
\label{p_r2tau50rho1delta0}
\end{figure}

Before considering explicit solutions, let us comment on the late time
behavior of the trajectories originating from a superposition state.
Provided the weight function $W(E,\tau_\textsc{b})$ has a compact
support in both its variables, the amplitude of the overall wave
function \eqref{GenBck} is asymptotically dominated by the component
with the smallest time development $q(\tau)$.  This can be seen by
considering the form of the single state solutions \eqref{GenStateX},
whose amplitude scales as
$|\psi_a(x,\tau)| \propto \exp\qty(-\frac{\xi x^2}{2 q_a^2})$, in the
case of $\nu = 1, \, n = 0$ as we consider here. Recalling that in the
large $|\tau|$ limit $q_a(\tau) \to q_\textsc{b} \omega_a \tau =
2 \sqrt{E_a} \tau$ --~see \eqref{qptau}~--, it follows that the component
with the smallest energy $E_a$ dominates the total wave
function \eqref{GenBck} in this regime.  Similarly, the total phase
$S$ resembles that of a single coherent state \eqref{Ssingle}, such
that the resulting trajectory is asymptotically classical; this
behavior is observed in all our results.  Since both the (rescaled)
quantum potential ($\tilde{Q}= Q/E_0$) and the semi-classical
potential ($\propto x^{-2}$), which become a function of time only
once evaluated along a trajectory, vanish asymptotically,
Eq.\,\eqref{fmaq} leads to
$x \underset{\tau\to\infty}{\propto} \tau$. Given that $x\propto
a^{\frac32 (1-w)}$ and $N\propto a^{3 w}$, this translates into
$a \underset{t\to\infty}{\propto} t^{2/[3(1+w)]}$, in which the cosmic
time $t$ is obtained from $N\dd\tau =\dd t$. In other words, also in
the two-state configuration \eqref{PhiB2}, the asymptotic behavior
follows the classical solution of general relativity for an FLRW
universe filled with a perfect fluid.  Furthermore, in the large
$|\tau|$ limit, the Hamilton-Jacobi equation reads $- \qty(x^2/\tau^2)
+ \dot{x}^2 = 0$, such that the trajectories satisfy
$\dot{x}(\tau)\large|_{|\tau| \to\infty} = \pm\, p_0$, where $p_0$ is
fixed by the initial condition $x(\tau_{\text{i}})$.  Hence, the
absolute value of a trajectory's momentum is equivalent at early and
late times, independent of the details of the quantum effects in
between.  For the sake of simplicity, and to guide the reader's eyes,
we always assume an initial condition for the trajectory $x (\tau)$
such that $x (\tau_\text{i}) = q_0(\tau_\text{i})$.

Fig.~\ref{x_r2tau50rho1delta0} shows an example of the kind of
trajectory one gets in our two-component state. We chose a situation
for which the two semi-classical states have different energies
($r=2$) and are quite widely separated in bouncing times, with
$\Delta\tau=50$, but enter the total wave function with equal
contributions ($\rho=1$).  The quantum trajectory, as expected from
the large-time behavior, first follows the coherent state
semi-classical trajectory selected by the initial condition $x
(\tau_\text{i}) = q_0(\tau_\text{i})$, bouncing for the first time
around $\tau\sim 0$, until it feels the effect of the second
state. For a finite range of times, this state dominates over the
first one, driving the evolution towards a second bounce at
$\tau \sim \Delta\tau$. After some transition time, the trajectory
ends up parallel to the reference one, again in agreement with the
expected behavior.

\begin{figure}[t]
    \includegraphics[scale=0.48]{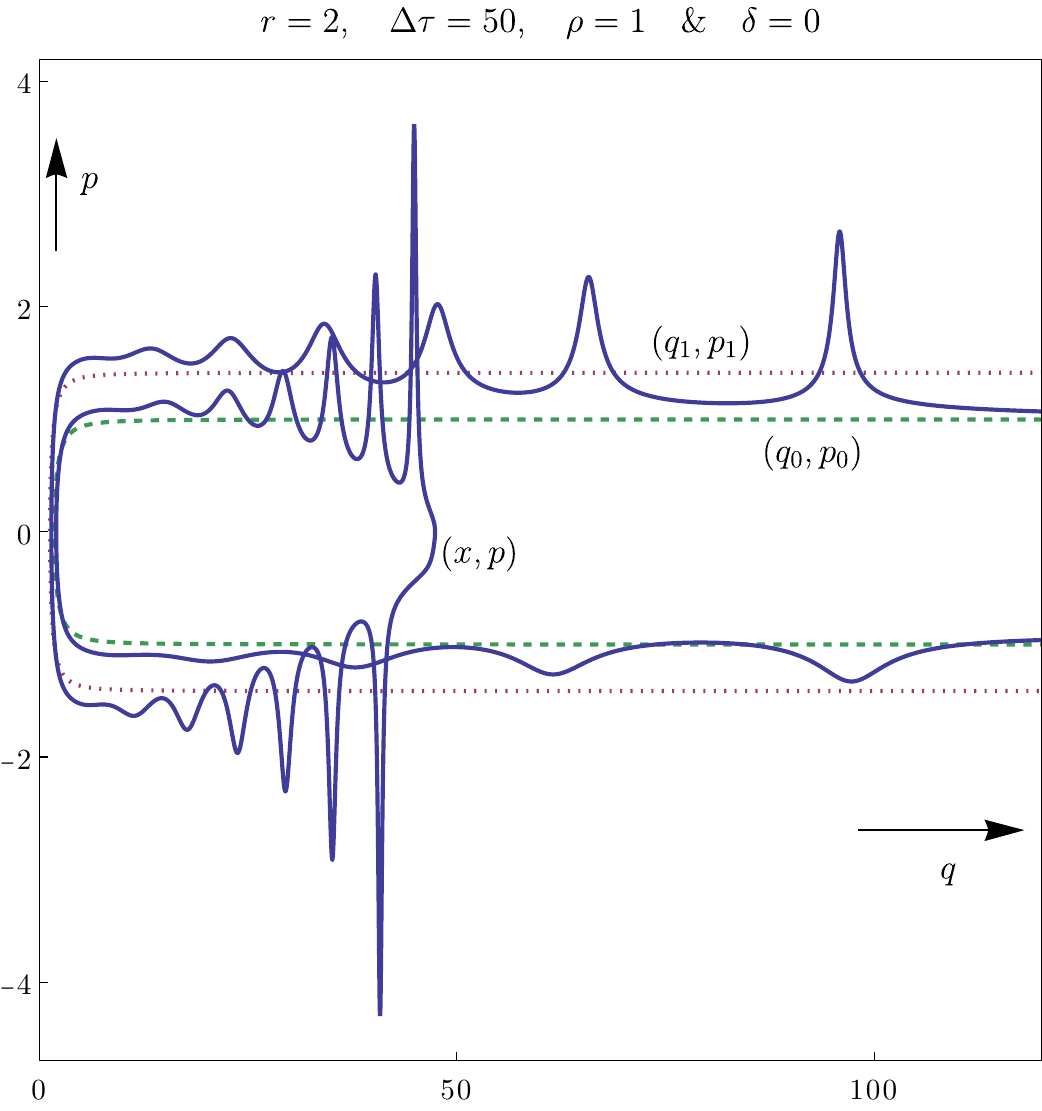}
    \caption{Phase space evolution of the trajectory for the
    two-component state \eqref{PhiB2} with the same
    parameters as in Figs.~\ref{x_r2tau50rho1delta0} 
    and \ref{p_r2tau50rho1delta0}, plotted with
    the same conventions. The quantum trajectory starts
    from the reference phase trajectory $(q_0,p_0)$,
    bounces once and subsequently connects to the second
    trajectory $(q_1,p_1)$ to bounce a second
    time before finally asymptotically getting back
    to the $(q_0,p_0$) trajectory.}
    \label{qp_r2tau50rho1delta0}
\end{figure}

In Fig.~\ref{x_r2tau50rho1delta0}, as apparent from the inset, the
trajectory exhibits regions of accelerations and decelerations that
are typical of quantum regimes; they are similar to those leading, to
the observation of an interference pattern with initially uniformly
distributed velocities in the two-slit experiment (see
e.g. \cite{Philippidis1979,holland1995quantum}). They appear most
clearly in Fig.~\ref{p_r2tau50rho1delta0}, which shows the
trajectory's momentum as functions of time, i.e. in our case, its
velocity $\dot{x}$. The acceleration and deceleration phases of
Fig.~\ref{x_r2tau50rho1delta0} are seen as sharp (but smooth) peaks in
the velocity. Combining both figures leads to the phase-space
representation of Fig.~\ref{qp_r2tau50rho1delta0}, showing that in the
mixed state, the quantum regime can be much more complicated than that
of each individual coherent state from which it is built. In
particular, although each basis state has a quantum regime lasting for
a given amount of time, that obtained from the combination can be much
larger. With perturbations generated during such a phase, one expects
such differences to be effectively measurable; we shall return to this
point in the following section.

\begin{figure}[t]
\includegraphics[scale=0.38]{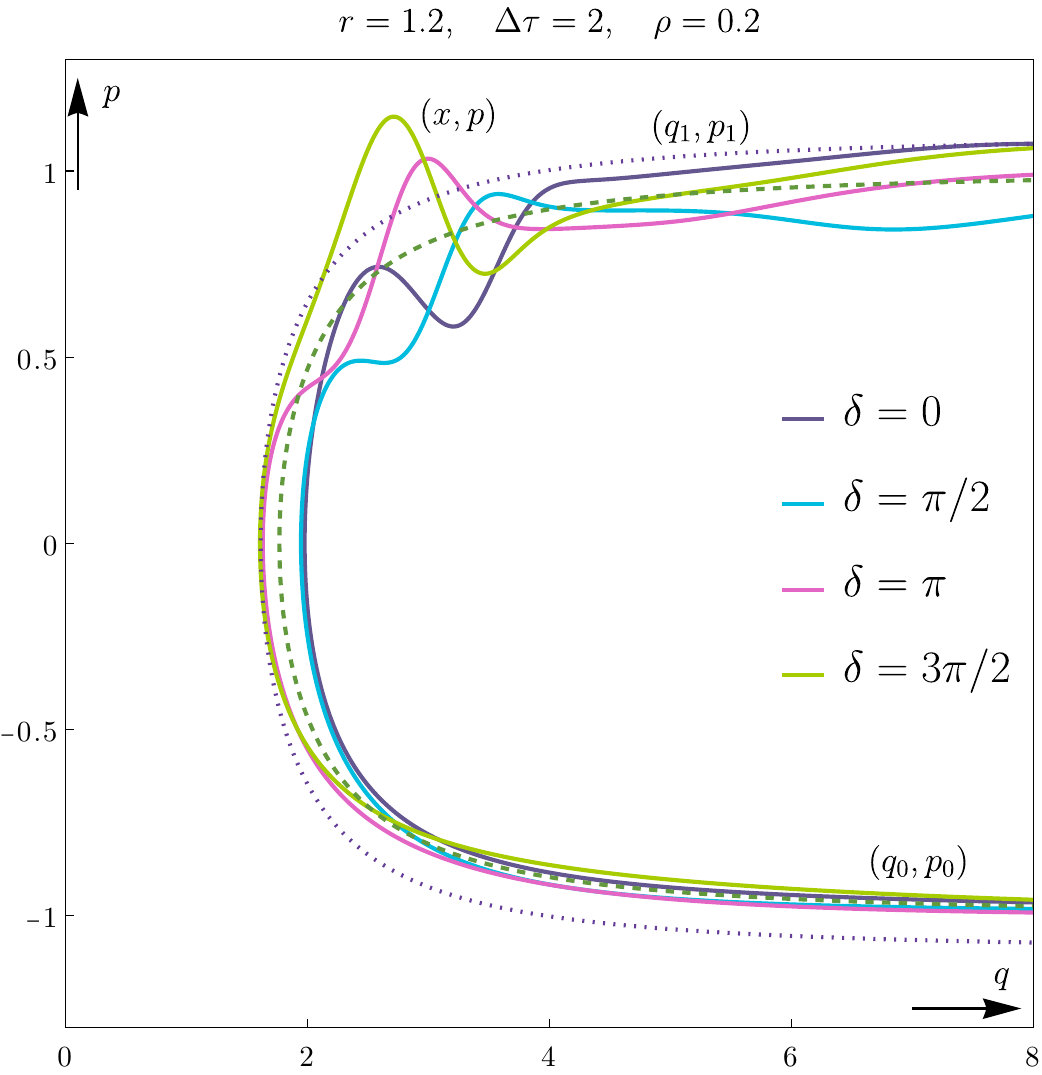}
\caption{Phase space evolution of multiple trajectories for the 
    two-component state \eqref{PhiB2} with a tiny energy difference
    $r=1.2$, small bouncing time delay $\Delta\tau=2$, small but not
    entirely negligible relative contribution $\rho=0.2$ of the second
    wave function, and different phases
    $\delta \in \qty{0,\frac{\pi}{2},\pi,\frac{3\pi}{2}}$.  This
    diagram shows that the relative phase $\delta$ impacts the quantum
    trajectory.  All trajectories return to their initial $|p|$ value
    at later times, which are outside the scope of this plot ($q \sim
    40$).  We set initial conditions at $\tau_{\text{i}} = -30$.}
\label{qp_r1.2tau2rho0.2deltaV}
\end{figure}

Fig.~\ref{qp_r1.2tau2rho0.2deltaV} depicts the phase space plots for
a different choice of parameters and different values of the phase
$\delta$ in \eqref{PhiB2}. In this case, the two states contributing
to the total wave function of the Universe are rather similar ($r =
1.2$ and $\Delta \tau = 2$), but the second one has only a relatively
small contribution with $\rho = 0.2$. Independent of the value of
$\delta$, the trajectories exhibit a single bounce that either follows
the subdominant semi-classical trajectory, or happens at a larger
scale than both trajectories. The oscillations in the phase space plot
already observed in Fig.~\ref{qp_r2tau50rho1delta0} are much smaller
in amplitude and contained to a smaller region. While all trajectories
have similar features, the specific value of $\delta$ clearly
influences the trajectories and can affect e.g. the bounce scale.

In general, the form of the trajectories depends strongly on the
choice of parameters, as well as the initial condition for the
trajectory.  For instance, for the parameters used in
Fig.~\ref{x_r2tau50rho1delta0}, an initial condition far away from a
semi-classical trajectory can lead to a single bounce only.  For
smaller $\Delta \tau$ the oscillations that can be found in
Fig.~\ref{qp_r2tau50rho1delta0} are generally smaller or absent and
the quantum effects are mostly limited to the bounce region.  On the
other hand, even a small contribution of a second wave function
$(\rho \sim 0.1)$ can lead to oscillations in $p$ for the resulting
trajectory, even away from the bounce, similar to what we saw in
Fig.~\ref{qp_r1.2tau2rho0.2deltaV}.  This is to say that there is a
plethora of possibilities for the trajectories that we show here and
the chosen examples are in that sense not generic.  However, and this
is the point we would like to emphasize, the trajectories we find
exhibit features that differ strongly from the semi-classical
solutions obtained for a single state wave function and thus introduce
generically different features in the Universe's evolution. For
instance, the full trajectory might include new length or time scales
stemming from the wave function that are absent from either wave
functions.  These features appear only due to the fact that we
consider trajectories.  We now turn to the possible impact of these
features on the evolution of perturbations.

\section{Tensor perturbations}
\label{sec:perturbations}

We left the concrete choice of fluid time general in the previous
sections, with the background defined by the variable $q\propto
a^{\frac32(1-w)}$ in terms of the scale factor $a$ depending on the
time $\tau$, itself.  The time $\tau$ is defined by the choice of the
lapse function $N\propto a^{3w}$.  The relation between the
trajectories and the scale factor as well as the lapse depend on the
actual fluid component driving the expansion of the universe in the
quantum regime through the equation of state parameter $w$. The regime
one is interested in in this work is that for which the universe is
sufficiently small such that quantum gravity effects may be relevant,
which usually requires a high temperature and thus a relativistic
fluid. In other words, if a fluid description is to be considered
adequate in the regime at hand here, the relevant fluid should be
radiation, so we can set $w\to \frac13$ in what follows. In this case,
one finds that $q\propto a$ and the fluid time is nothing but the
conformal time $\tau\to \eta$.

\subsection{Born-Oppenheimer mode evolution}

As is usual in a trajectory approach such as the one discussed
above~\cite{Pinto-Neto:2013npa,Pinto-Neto:2013toa,Peter:2015zaa}, we
consider a Born-Oppenheimer approximation (see, e.g., Refs.~\cite{Kiefer:2025udf,Bergeron:2025eda}) to describe the quantum
evolution of primordial perturbations. We restrict our attention to
the tensor modes (gravitational waves), as those can be seen as
spectator fields on the evolving
background~\cite{Micheli:2022tld}. Setting the full metric to take the
form
\begin{align}
\dd s^2 = a^2(\eta) \qty{ -\dd \eta^2 + \qty[\delta_{ij}
+ h_{ij} \qty(\bm{x},\eta)] } \dd x^i\dd x^j,
\label{ds2hij}
\end{align}
with transverse ($\partial^i h_{ij} =0$) and traceless ($\delta^{ij}
h_{ij} = 0$) tensor perturbations $h_{ij}$, we can expand the
Einstein-Hilbert action to second order as
\begin{equation}
\delta^{(2)} \mathcal{S}_\textsc{eh} = \dfrac{1}{8\kappa}
\int\dd[4]{x} a^2(\eta) \qty( \pdv{h^i_{\ j}}{\eta} \pdv{h^j_{\ i}}{\eta}- \partial_k h^i_{\ j} \partial^k h^j_{\ i}),
\label{delta2T}
\end{equation}
which, upon expanding into discrete spatial modes (recall that one considers
compact spatial hypersurfaces with finite volume $\mathcal{V}_0$)
\begin{equation}
h_{ij}\qty(\bm{x},\eta) = \sqrt{\dfrac{4\kappa}{\mathcal{V}_0}} \sum_{\bm{n},\lambda} \varepsilon_{ij}^{(\lambda)}
\dfrac{\mu^{(\lambda)} \qty(\bm{n},\eta)}{a(\eta)}
e^{2\ii \pi \bm{n}\cdot\bm{x}/L},
\label{h2mu}
\end{equation}
with $\bm{n} = \qty(n_x,n_y,n_z)$ an integer-valued vector and
$L=\mathcal{V}_0^{1/3}$ the spatial extension in each direction 
(we assume a torus topology for 3-space). Eq.~\eqref{h2mu}
defines the helicity modes $\mu^{(\lambda)}$ along the polarizations
$\varepsilon_{ij(\pm)}$ as in Ref.~\cite{Micheli:2022tld}
with the extra complication of compact spatial sections. 
Defining the comoving wavevector $\bm{k} = 2\pi\bm{n}/L$, 
the relation above can be put in the form $h_{ij} = \sum_{\bm{n},\lambda}\varepsilon_{ij}^{(\lambda)}e^{\ii\bm{k}\cdot\bm{x}} h^{(\lambda)}_k$, i.e. $a h^{(\lambda)}_k =  \sqrt{4\kappa/\mathcal{V}_0} \mu^{(\lambda)}_k$ and one gets
the second-order Hamiltonian $H^{(2)}
= \sum_{\bm{k}} \qty[ H^{(2)}_{\bm{k},+} + H^{(2)}_{\bm{k},-}]$,
with
\begin{align}
H^{(2)}_{\bm{k},\lambda} = 
\pi^{(\lambda)}_{\bm{k}} \pi^{(\lambda)}_{-\bm{k}}
+ \left( k^2 - \frac{a''}{a} \right)
\mu^{(\lambda)}_{\bm{k}} \mu^{(\lambda)}_{-\bm{k}}.
\label{H2kini}
\end{align}
In Eq.~\eqref{H2kini}, the sum is over wavevectors pointing upwards
only, $\mu^{(\lambda)}_{\bm{k}}$ is a Fourier mode of
$\mu^{(\lambda)}$, with conjugate momentum $\pi^{(\lambda)}_{\bm{k}}$,
and a prime denotes derivative with respect to the conformal time
$\eta$, $a' = \dd a /\dd\eta$. Both types of helicity
modes follow the same dynamics, so we drop the superscript $(\lambda)$
in what follows.

Upon quantization of \eqref{H2kini}, which is also detailed in
Ref.~\cite{Micheli:2022tld}, one can extract the gravitational wave
power spectrum through the evaluation of mode functions
$\mu_{\bm{k}}(\eta)$, usually setting vacuum initial conditions. For a
given comoving wavenumber $k = \sqrt{\bm{k}^2}$, the
Hamiltonian \eqref{H2kini} provides the mode equation
\begin{align}
\mu_{\bm{k}}'' + \qty[ k^2  - V_\text{eff}(\eta)] \mu_{\bm{k}} = 0\,,
\label{EqMode}
\end{align}
in which the effective potential 
\begin{align}
V_\text{eff}(\eta) \equiv \frac{a''}{a}
\label{Veff}
\end{align}
provided by the time-dependent background acts as a source of quantum
excitations for the tensor perturbation modes.  Note that even though
we restrict our attention here to gravitational waves, a similar
equation also holds for the scalar modes responsible for the major
part of the temperature fluctuations measured in the cosmic microwave
background~\cite{Mukhanov:1990me}.

\begin{figure}[t]
\includegraphics[scale=0.41]{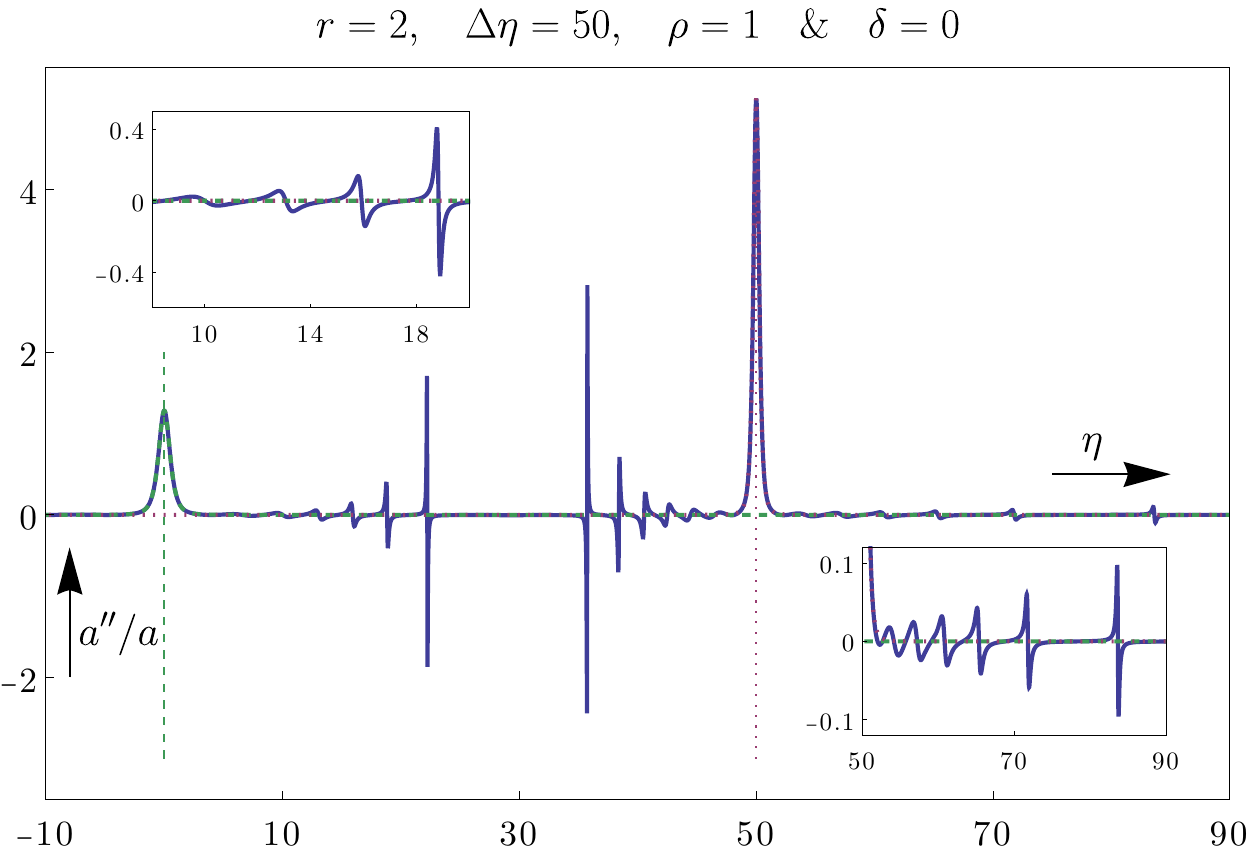}
\caption{Effective potential \eqref{Veff} needed to solve the mode
    equation \eqref{EqMode} obtained from the trajectory of
    Fig.~\ref{x_r2tau50rho1delta0} and the same underlying parameters.
    The individual effective potentials, calculated through
    $q_0''/q_0$ and $q_1''/q_1$, are shown as dashed and dotted lines,
    respectively. Additionally, the bounce times are respectively
    marked at locations $\tau=0$ and $\tau=50$.}
\label{Veff_r2tau50rho1delta0}
\end{figure}

\begin{figure}[t]
\includegraphics[scale=0.49]{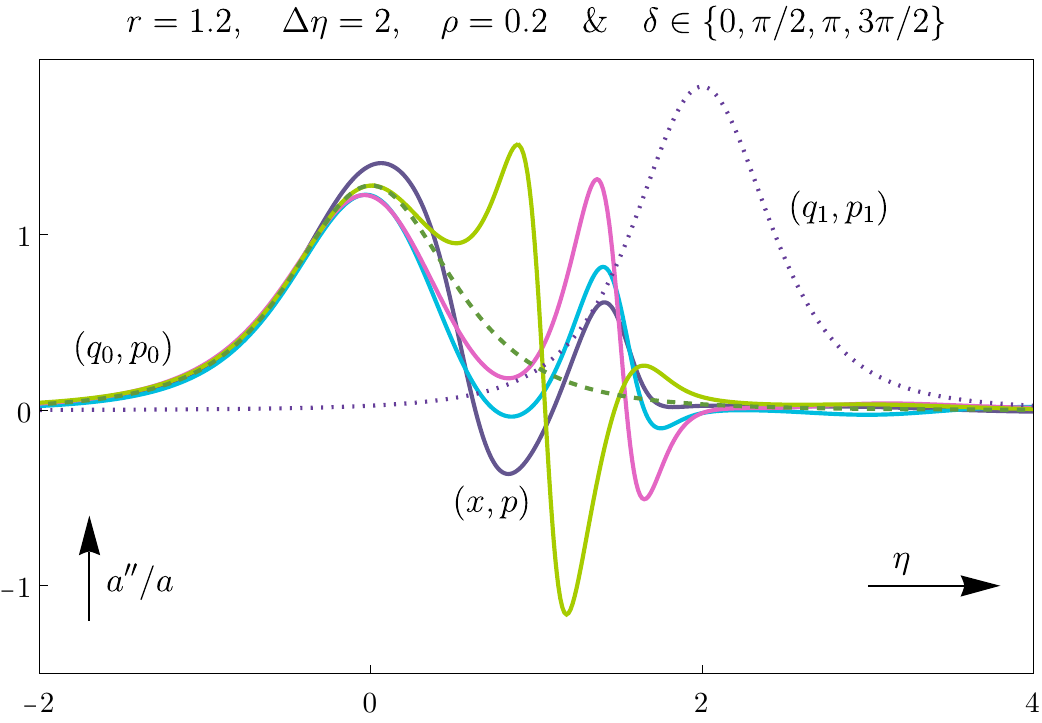}
\caption{Effective potential \eqref{Veff} needed to solve the mode
    equation \eqref{EqMode} obtained from the trajectories of
    Fig.~\ref{qp_r1.2tau2rho0.2deltaV} (same underlying parameters).
    The individual effective potentials calculated from $q_0''/q_0$
    and $q_1''/q_1$ are shown as dashed and dotted lines,
    respectively. These peak at their respective bounce times.  The
    full lines represent the same values of the relative phase
    $\delta$ as in Fig.~\ref{qp_r1.2tau2rho0.2deltaV}.}
\label{Veff_r1.2tau2rho0.2deltaV}
\end{figure}

Fig.~\ref{Veff_r2tau50rho1delta0} and \ref{Veff_r1.2tau2rho0.2deltaV}
present two examples of $V_\text{eff}(\eta)$,
with the same choices of parameters as for the trajectories shown in
Fig.~\ref{qp_r2tau50rho1delta0} and \ref{qp_r1.2tau2rho0.2deltaV},
respectively.  These figures demonstrate that in the trajectory
approach to quantum cosmology, the effective potential of a
superposition state can exhibit widely different features to the
single state case while being heavily dependent on the chosen
parameters in the biverse case.

In Fig.~\ref{Veff_r2tau50rho1delta0}, which depicts the effective
potential resulting from the trajectory with widely separated bounces
shown in Fig.~\ref{x_r2tau50rho1delta0}. Perturbations are produced
at both bounce times (we write $\Delta\eta$ instead of $\Delta\tau$),
with $V_\text{eff}(\eta)
\underset{\eta\sim0}{\simeq} q_0''/q_0$ and $V_\text{eff}(\eta)
\underset{\eta\sim \Delta\eta}{\simeq} q_1''/q_1$, 
as well as in between the bounces and after the second bounce, where
$V_\text{eff}(\eta)$ exhibits stark fluctuations.  This induces
potentially large contributions to the production of perturbations for
a longer time in comparison with the single state case. Since the
additional fluctuations in the potential are highly peaked, with both
positive and negative values, one might approximate these
contributions by Dirac $\delta$-like potentials. They lead to serious
modifications of the power spectrum~\cite{Martin:2003bp} produced by
either $q_0''/q_0$ or $q_1''/q_1$ alone, not to mention that merely
adding these two contributions ($q_0''/q_0$ and $q_1''/q_1$) already
changes the predicted spectrum, including, e.g., superimposed
oscillations~\cite{Falciano:2008gt}.  In this particular example, the
effective potential follows the contributions of both single wave
function components due to the fact that the relative amplitude
contribution $\rho$ has been set to unity.

Fig.~\ref{Veff_r1.2tau2rho0.2deltaV} shows the effective potential of
the perhaps less spectacular case of two rather similar bounces with
comparable energies and for different values of the phase $\delta$,
the trajectories of which are depicted in
Fig.~\ref{qp_r1.2tau2rho0.2deltaV}.  In this case, the full effective
potential is dominated by the first bounce, as expected from the
choice $\rho=0.2$.  Depending on the relative phase $\delta$, however,
the effective potential is altered, exhibiting fluctuations that are
less peaked but can similarly be negative.  Once again, one expects
the power spectrum produced in such a model to have very different
characteristic features from the single state case, depending on the
actual trajectory chosen, i.e. on the initial condition for the scale
factor when the universe enters the quantum gravity regime, but also
on the parameters of the quantum state.

These examples illustrate that the non-standard behavior of
trajectories translates into the dynamics of perturbations and one can
therefore generally expect noticeable effects.  This may have serious
consequences for the cosmological predictions depending on which
version of quantum mechanics is used to evaluate them.

In the standard Copenhagen-inspired view, the universe follows one of
the coherent states in the superposition, say $\ket{q_0(\eta),
p_0(\eta)}$, and thereby traces the semi-classical trajectory,
$q_0(\eta)$.  The calculation of the cosmological perturbations is
done in each branch of the full wave function independently, leading
to a spectrum which is rather similar in the different branches since
$q_0''/q_0$ and $q_1''/q_1$ have very similar behaviors. The final
step consists in projecting onto the relevant state.  If one relaxes
the Born-Oppenheimer assumption, one finds that the existence of the
other state $\ket{q_1(\eta), p_1(\eta)}$ can lead to corrections in
the primordial power spectrum stemming from entanglement between the
perturbations of both
components~\cite{Bergeron:2024art,Bergeron:2025eda}.  While the
predicted spectrum remains almost unchanged in comparison to the
single state case, the effect of the superposition state appears
through the production of primordial non-gaussianities, which result
from the coupling between the different background components.

In the trajectory approach, however, the situation is entirely
different.  There is no projection involved at any step of the
calculation, as one assumes the quantum trajectory to represent the
actual motion of the scale factor as a function of time.
Specifically, there exists an actual trajectory $a(\eta)$ at all times
$\eta$, and this trajectory is given by the time development of the
total background wave function.  The perturbations then propagate on
the background defined by the trajectory resulting from the total wave
function.  This is in contrast to the formalism developed in
Ref.~\cite{Bergeron:2025eda}, where one considers the evolution of
perturbations within different background states that influence each
other only once the approximation of a Born-Oppenheimer factorization
is broken.

To obtain the dynamics of perturbations one needs to calculate the
full background evolution first, which can be influenced by several
components of the background Hilbert space and is described by a fully
quantum trajectory. Contrary to the usual lore, the background state
has no reason to be well localized around some semi-classical
trajectory from which one could approximate the asymptotically
classical Universe we end up living in: in the present case, what
matters is that the asymptotic trajectory evolves towards an almost
classical one, which is ensured by demanding that the quantum
potential decays for sufficiently large values of the scale factor.

Thus, one finds that the two formalisms, although obviously completely
equivalent in a statistical setup, i.e., if one can repeat the process
a large number of times, may lead to radically different predictions
in the cosmological case, where there exists only a single
`experiment' for the evolution of the background on which
perturbations propagate.

Indeed, examining Fig.~\ref{Veff_r2tau50rho1delta0}
and \ref{Veff_r1.2tau2rho0.2deltaV} reveals that the structure of the
resulting particle production effective potential $V_\text{eff}(\eta)
= a''/a$ can be widely different from either $q_0''/q_0$ and
$q_1''/q_1$.

\subsection{Calculating the power spectrum}

In order to assess the influence of the trajectories on the power spectrum of tensor perturbations, we compare two power spectra obtained from biverse trajectories to that of the single state case. 
The tensor power spectrum~\cite{Peter:2013avv} $\mathcal{P}_h$
can be calculated using the relationship~\eqref{h2mu} between
$h_k$ and $\mu_k$, leading to (counting both helicities)
\begin{align}
    \mathcal{P}_h(k, \eta) = \dfrac{k^3}{2 \pi^2} 2 |h_k|^2= 
    \dfrac{48 k^3}{\pi^2} \left|\dfrac{\mu_k(\eta)}{q(\eta)}\right|^2, \label{Phk}
\end{align}
where we set $q(\eta)=\sqrt{12\mathcal{V}_0/\kappa}a(\eta)$
in agreement with Eq.~\eqref{qanda} for $w=\frac13$. Note
that in this special case, the effective potential of \eqref{Veff}
can be replaced by $q''/q$ and, the resulting
spectrum \eqref{Phk} depends on the ratio $\mu/q$.
The single state case, for which the effective potential is obtained from the semiclassical solution \eqref{qptau}, has been studied in detail in \cite{Peter_2006}. 
We would like to point out that since the one state time-dependent expectation value and the trajectory obtained in Eq.~\eqref{xdBBSingle} are proportional, the effective potential in this case is the same for the trajectory and Copenhagen approaches to quantum mechanics.

The integral version of the perturbation equation of motion \eqref{EqMode} reads \cite{Mukhanov:1990me}
\begin{align}
\begin{split}
    \dfrac{\mu_k (\eta)}{q(\eta)} =& \frac{\mu_{k,\text{i}}}{q_\text{i}} + (\mu_{k,\text{i}}' q_\text{i} - \mu_{k,\text{i}} q_\text{i}') \int_{\eta_\text{i}}^\eta \frac{\dd \tilde{\eta}}{q^2(\tilde{\eta})} \\
    & \qquad - k^2 \int_{\eta_\text{i}}^\eta \frac{\dd \tilde{\eta}}{q^2(\tilde{\eta})} \int_{\eta_\text{i}}^{\tilde{\eta}} q(\tilde{\tilde{\eta}}) \mu_k (\tilde{\tilde{\eta}}) \dd \tilde{\tilde{\eta}} ,
\end{split}
\label{eq:muEeomInt}
\end{align}
where the subscript 'i' indicates an initial condition set at some initial time $\eta_\text{i}$, i.e. $q_\text{i} \equiv q(\eta_\text{i})$, etc. 
We set initial conditions for the perturbations in the far pre-bounce phase at a time for which $k^2\gg q''/q$ 
as the
Bunch-Davies vacuum, namely $\mu_k = \frac{1}{\sqrt{2 k}}e^{- \mathrm{i} k (\eta - \eta_\text{i})}$~\cite{Peter:2013avv}, such that $\mu_{k, \text{i}} \equiv \mu_k(\eta_\text{i}) = 1/\sqrt{2k}$ and $\mu_{k,\text{i}}'  = -\ii \sqrt{k/2}$~\cite{Mukhanov:1990me}.

In the 
regime in which the effective potential dominates, i.e. $|V_{\text{eff}}|\gg k^2$,
one can focus on the first two terms of Eq.~\eqref{eq:muEeomInt} to get two modes for $\mu_k/q$, namely a constant and a term proportional
to $\arctan [\omega(\eta-\eta_\textsc{b})]$. 
While evolving
from $\eta<\eta_\textsc{b}$ to $\eta>\eta_\textsc{b}$, the latter goes from
one plateau to another, so the tensor perturbation 
is a constant outside the bounce regime.

Finally, the effective potential becomes subdominant again 
and the solutions for $\mu_k$ correspond to plane waves. Applying these approximations to a generic constant equation of state leads to a tensor spectral index $n_\textsc{t} = 12 w/(1+3w)$~\cite{Peter_2006}, i.e. $\mathcal{P}_h \propto k^2$ for the case of radiation.

To calculate the power spectrum from numerical solutions for the $\mu_k$, we consider a certain range of $k-$modes, namely $k \in [k_{\mathrm{min}}, k_{\mathrm{max}}] = [3.\times 10^{-4}, 8.\times 10^{-3}]$. The power spectrum $\mP_h$ is calculated at a time $\eta^*$, where $|V_{\text{eff}}| \geq k^2$ for all $k-$modes in question, specifically, we choose $\eta^*$ such that $|V_{\text{eff}}(\eta^*)| = k_{\mathrm{max}}^2$.

The range of scales chosen may seem somehow arbitrary. 
The connection between
the numerical values of the comoving wavenumber $k$ and the
physical wavelength $\lambda_\text{phys} = 2\pi a_\text{now}/k$
is not immediate as it requires knowledge of the value of the
scale factor today $a_\text{now}$, which demands that the full
history of the Universe be known. 
For large
wavelengths ($k<k_\text{min}$) and in all cases studied, one has
reached the asymptotic limit so it is unnecessary to go below that. 
Small scales are either unchanged for $k>k_\text{max}$ (case of
a Universe dominated by one state) or the potentially relevant effects
discussed below are already clearly visible.

\begin{figure}[t]
\includegraphics[width = 0.45\textwidth]{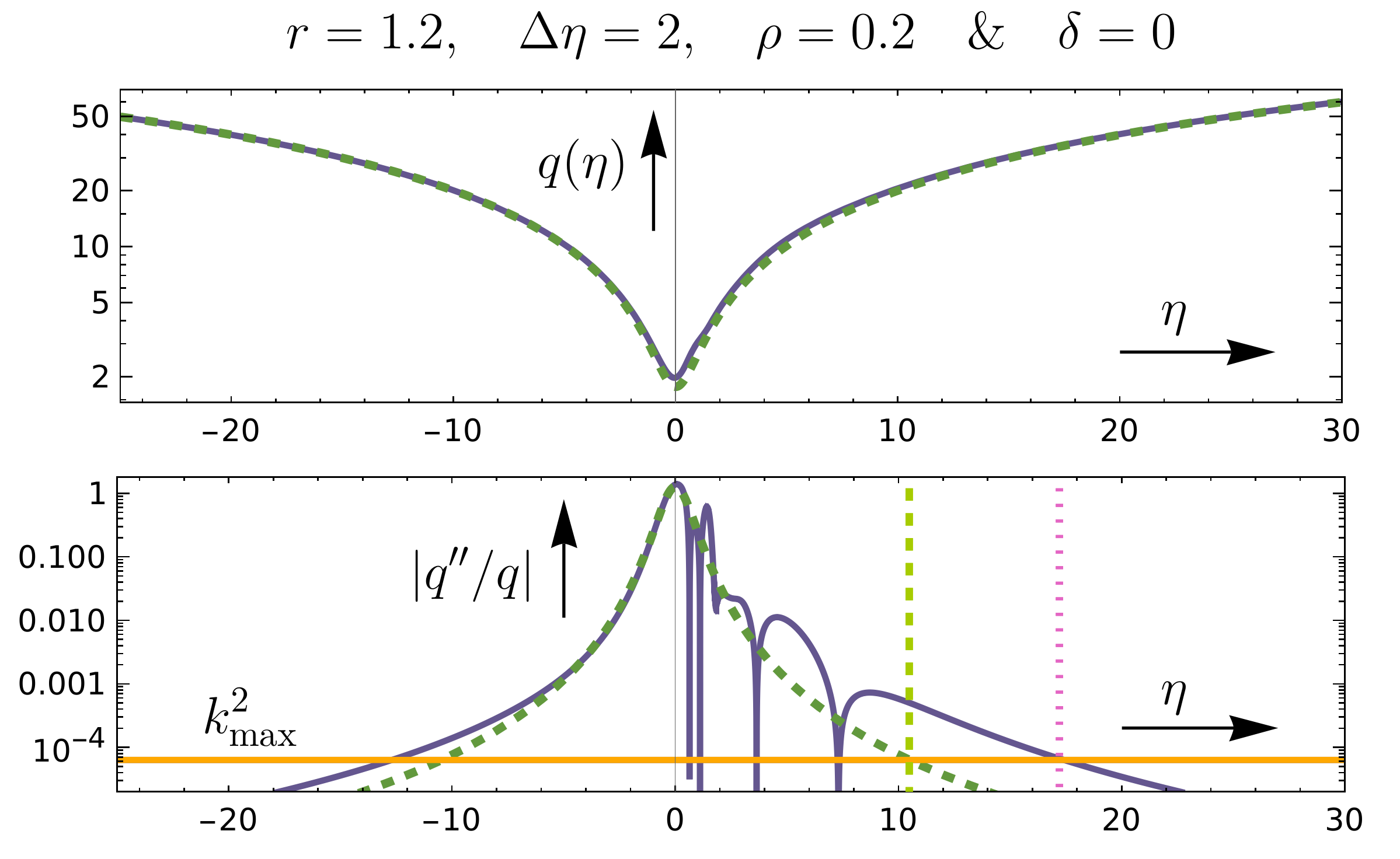}
\caption{ 
\emph{Top panel:} The bouncing trajectories for the biverse (purple, full) as well as for the single state case (green, dashed). 
\emph{Bottom panel:} The absolute value of the effective potential $|V_{\mathrm{eff}}|$ for the biverse case (purple, full)  as well as for the single semiclassical trajectory $q_0$ (green, dashed), which are also depicted in Fig.~\ref{Veff_r1.2tau2rho0.2deltaV} for $\delta = 0$. The horizontal orange line corresponds to $k_{\mathrm{max}}^2 = 6.4\times 10^{-5}$. 
The vertical lines represent the times $\eta^*$ at which the perturbative dynamics cease to be dominated by the effective potential (for the last time) for the single state case (green, dashed) and the biverse case (pink, dotted). 
Note that some of the peaks in $|V_{\mathrm{eff, \, biverse}}|$ correspond to oscillations (cf. Fig.~\ref{Veff_r1.2tau2rho0.2deltaV}). 
}
\label{fig:delta0_Veff_kMax}
\end{figure}

\begin{figure}[t]
\includegraphics[width=0.48\textwidth]{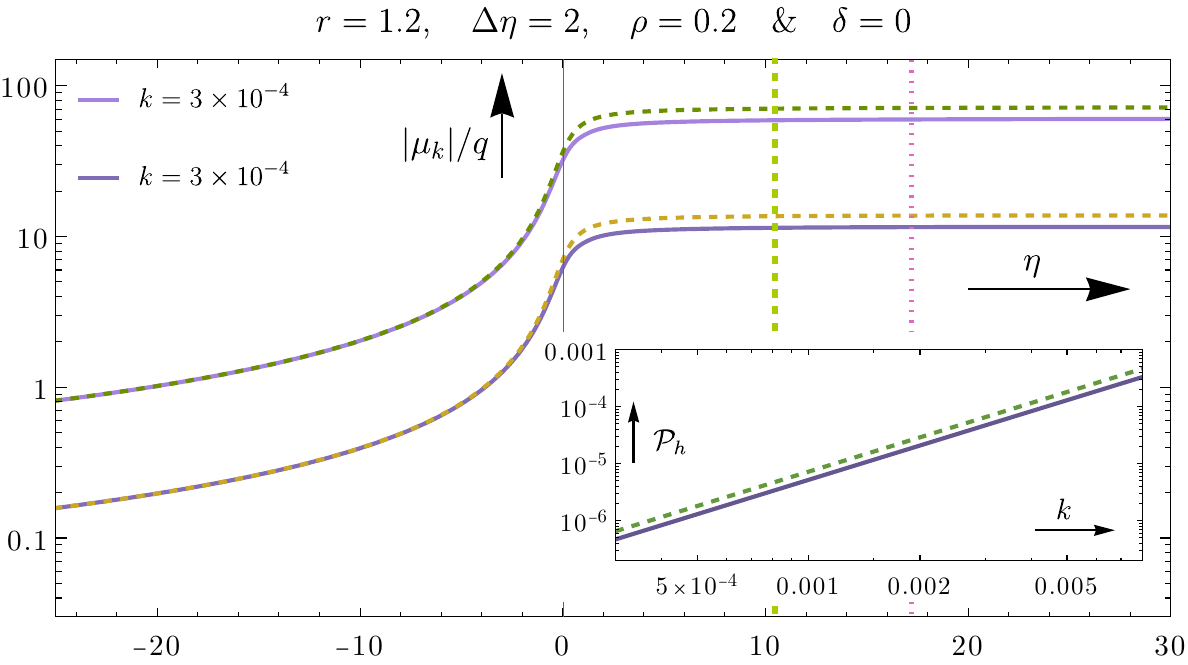}
\caption{
The evolution of $|h_k(\eta)|$ for $k_{\mathrm{min}}$ and $k_{\mathrm{max}}$. The full lines correspond to the evolution obtained for the biverse trajectory with $\delta = 0$ shown in Fig.~\ref{qp_r1.2tau2rho0.2deltaV} 
and Fig.~\ref{fig:delta0_Veff_kMax}, and the dashed lines give the corresponding evolution of $|h_k|$ for the single state trajectory $q_0$ (shown in the same plots as the biverse trajectory). 
The bounce occurs at $\eta = 0$.
The vertical lines indicate the time $\eta^*$ at which the power spectrum $\mP_k$ is calculated (green, dashed for the single state case and pink, dotted for the biverse case), which is chosen such that $|V_{\text{eff}}(\eta^*)| = k_{\mathrm{max}}^2$.
The inset shows the tensor power spectrum for a background given by the biverse trajectory (full, purple), as well as for the single state case (green, dashed).
While there is an overall decrease in power in the biverse case, the spectral index is almost the same as in the single state case (up to the third significant figure in a best fit). 
}
\label{fig:delta0_hk}
\end{figure}

For the biverse trajectories, let us first consider an example where the second wave function enters the total state with a relatively small weight $\rho = 0.2$ and the bounces occur fairly close to one another $\Delta \tau = \Delta \eta = 2$, namely we consider the trajectory depicted in Fig.~\ref{qp_r1.2tau2rho0.2deltaV} with $\delta = 0$. 
Fig.~\ref{fig:delta0_Veff_kMax} shows the biverse and single state case trajectory as a function of time, as well as $k_{\mathrm{max}}^2$ in comparison to the absolute values of the effective potentials $|V_{\mathrm{eff}}|$. 
The biverse trajectory differs only slightly from the single state case around the bounce region. 
As can be seen in the bottom panel of Fig.~\ref{fig:delta0_Veff_kMax}, the perturbations evolving on the biverse trajectory are influenced by the effective potential for longer ($\eta* \approx 17.2$ instead of $\eta^* \approx 10.5$). 
The evolution of perturbation modes $|h_k|$ for $k= k_{\mathrm{min}}$ and  $k= k_{\mathrm{max}}$ are shown in Fig.~\ref{fig:delta0_hk}. 
At the bounce, which influences the perturbations on the biverse as well as the single state trajectory, the absolute value of perturbations increases, although it should be noted that this increase is smaller in the biverse case. The difference can be attributed to the fact that
the 
initial
crossing point $k^2 = |V_\text{eff}|$ in the contracting branch happens sooner for the
biverse case than for the single state (Fig.~\ref{fig:delta0_Veff_kMax})
leading 
to an initially 
lower value for $|\mu_k/a|$.
The perturbations remain approximately constant after the bounce in both cases, again as expected. 

The resulting power spectra for both cases are shown as an inset in Fig.~\ref{fig:delta0_hk}.
In the single state case we find $\mP_h \sim k^2$, consistent with the result of Ref.~\cite{Peter_2006}. 
The biverse spectrum has almost, i.e. up to numerical accuracy, the same spectral index as the single state case for the chosen range of $k-$values with an overall slightly smaller amplitude.


\begin{figure}[t]
\includegraphics[width=0.48\textwidth]{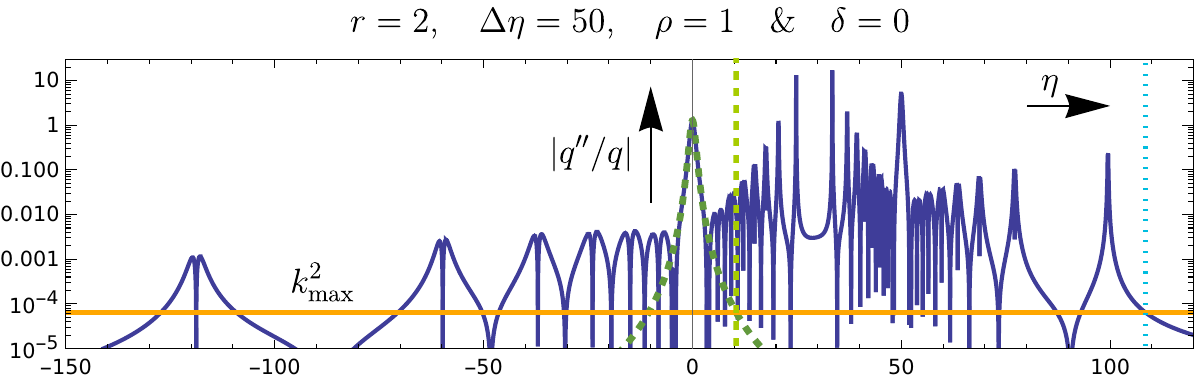}
\caption{ 
The absolute value of the effective potential $|V_{\mathrm{eff}}|$ for the biverse case (blue, full)  as well as for the single semiclassical trajectory $q_0$ (green, dashed), which are also depicted in Fig.~\ref{Veff_r2tau50rho1delta0}. The horizontal orange line corresponds to $k_\text{max}^2 = 6.4\times 10^{-5}$. We thus see that the effects of the effective potential become relevant much sooner and affect the evolution of the perturbations for much longer than in the single state case. 
The vertical lines represent the times $\eta^*$ at which the perturbative dynamics cease to be dominated by the effective potential (for the last time) for the single state case (green, dashed) and the biverse case (blue, dotted). 
Note that most of the peaks in $|V_{\mathrm{eff, \, biverse}}|$ correspond to oscillations (cf. Fig.~\ref{Veff_r2tau50rho1delta0}) and only the regions around the two bounces at $\eta = 0$ and $\eta = 50$ are unaccompanied by a negative peak. 
}
\label{fig:deltaTau_Veff_kMax}
\end{figure}

\begin{figure}[t]
\includegraphics[width=0.48\textwidth]{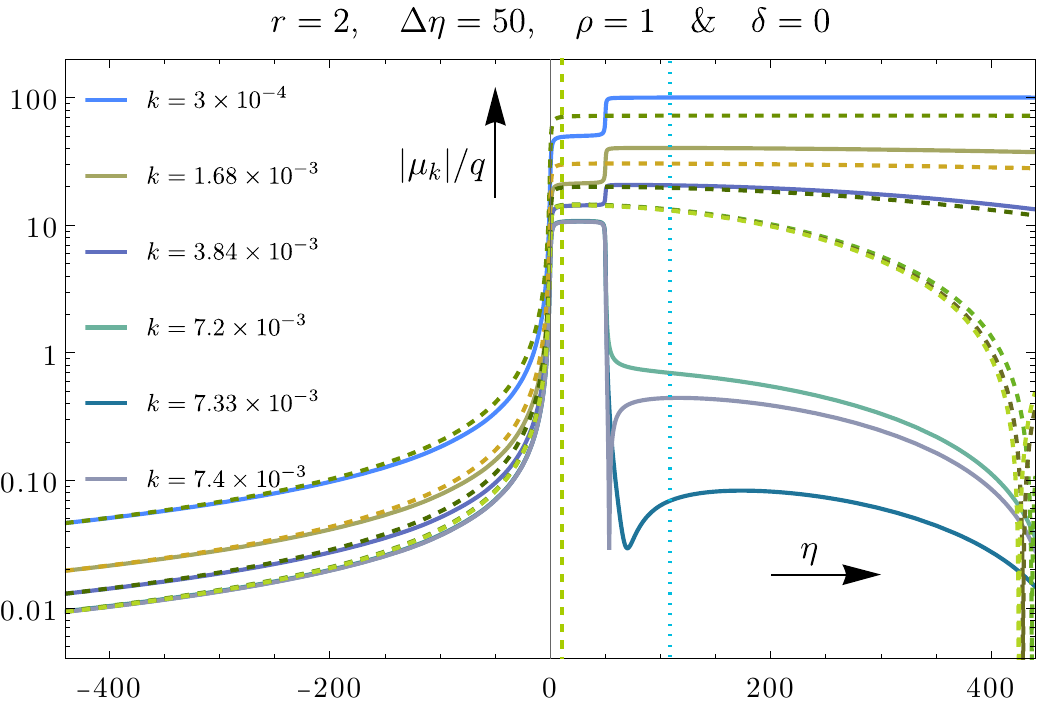}
\caption{
The evolution of $|h_k(\eta)|$ for different values of the wavenumber $k$, as indicated in the plot. The full lines correspond to the evolution obtained for the biverse trajectory shown in Fig.~\ref{x_r2tau50rho1delta0} with the effective potential depicted in Fig.~\ref{Veff_r2tau50rho1delta0} and Fig.~\ref{fig:deltaTau_Veff_kMax}, and the dashed lines give the corresponding evolution of $|h_k|$ for the single state trajectory $q_0$ (shown in the same plots as the biverse trajectory). 
The first bounce, experienced by the single state and biverse case, occurs at $\eta = 0$, whereas the second influences only the perturbations in the biverse and takes place at $\eta = 50$.
The vertical lines indicate the moment at which the power spectrum $\mP_k$ is calculated (green, dashed for the single state case and blue, dotted for the biverse case). 
}
\label{fig:deltaTau_hk}
\end{figure}

As a second example, we consider the spectrum obtained for the biverse trajectory depicted in Fig.~\ref{x_r2tau50rho1delta0} where the two states of the superposition have a large separation in bounce times $\Delta \eta = 50$ and enter the total wave function with equal weights $\rho = 1$. 
Fig.~\ref{fig:deltaTau_Veff_kMax} shows the absolute values of the effective potentials for the biverse and the single state case in relation to $k_{\mathrm{max}}^2$, indicating the regimes in which the perturbative dynamics \eqref{EqMode} are dominated by $V_{\mathrm{eff}}$. 
In comparison with the single state case, the effective potential of the biverse influences the perturbations much earlier and for much longer; the largest wavenumber in the range we consider here exits the regime of potential domination at $\eta^* \approx 108.5$ instead of $\eta^*\approx 10.5$ in the single state case. 
The evolution of several perturbation modes $|h_k|$ are shown in Fig.~\ref{fig:deltaTau_hk}. 
They exhibit the same behaviour as found in the previous example around the  time of the first bounce, namely an overall increase in perturbation amplitude in accordance with Ref.~\cite{Peter_2006}, again with slightly lower amplitude than the single bounce case.
For the single trajectory case, perturbations exit the potential dominated regime shortly after the bounce; for larger $k-$modes the first oscillation of the perturbations can be seen in the plot. 

Perturbations propagating on a biverse trajectory on the other hand are influenced by the second bounce and experience another shift in amplitude, whose sign depends on the regime in $k$. Indeed, it is interesting to note that for small $k$, the amplitude is increased
again (second plateau in Fig.~\ref{fig:deltaTau_hk}), thereby leading to an overall larger amplitude, while larger $k$ experience a drop
after the first increase, with a resulting amplitude less than
that of the single bounce case.

Keeping in mind the solution for perturbations in a potential dominated phase, namely Eq.~\eqref{eq:muEeomInt} without the last term, the
sign of the second shift in amplitude can be understood in that
$|h_k|$ depends on the initial conditions of the respective mode upon encountering the second bounce regime.
To understand why the main effect of the biverse trajectory is rooted in the second bounce and the many oscillations of $V_{\mathrm{eff}}$ that can be seen in Fig.~\ref{Veff_r2tau50rho1delta0} appear to have little effect, it is instructive to recall the form of the background trajectory (Fig.~\ref{x_r2tau50rho1delta0}) and that the dynamics are governed by 
$\int_{\eta_{\rm i}}^\eta q^{-2}(\tilde{\eta})\dd{\tilde{\eta}}$:
the oscillations of $V_{\mathrm{eff}}$ correspond to small fluctuations in 
$q(\eta)$ 
that only slightly deviate from the semiclassical trajectories of the two states in the superposition, so that the trajectory approximately follows the semiclassical solutions. 

\begin{figure}[t]
\includegraphics[width=0.48\textwidth]{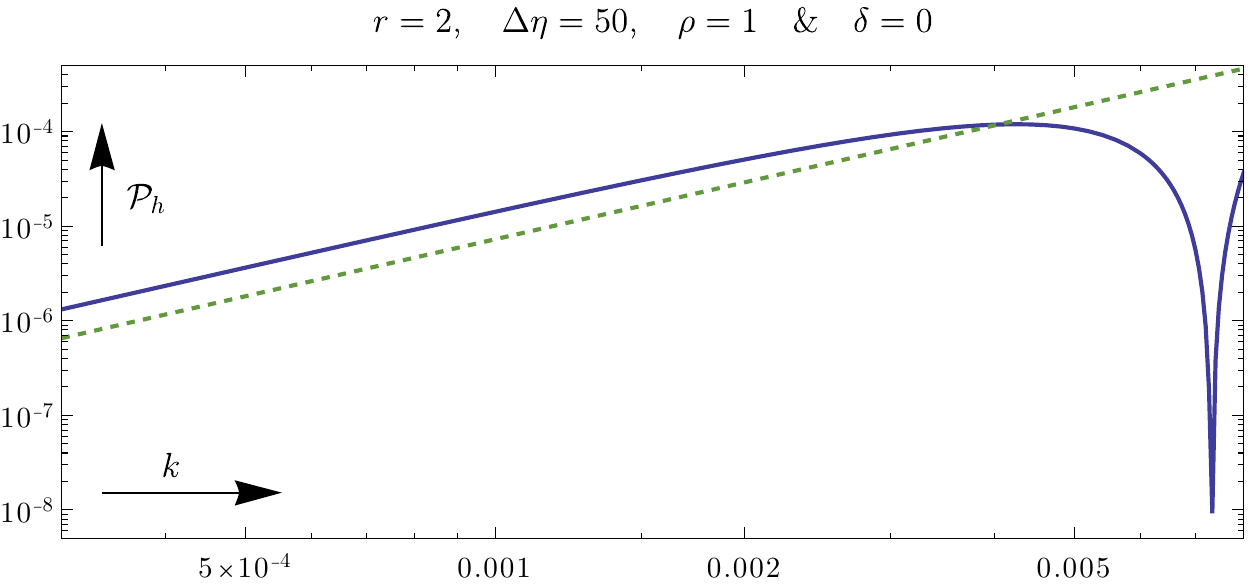}
\caption{
Tensor power spectrum for a background given by the biverse trajectory (blue, full) depicted in Fig.~\ref{x_r2tau50rho1delta0}, as well as for the single state case $q_0$ (green, dashed).
For large wavelengths (small $k$), the spectrum is roughly unchanged (up to numerical errors) with a slight increase in amplitude. 
For smaller scales (large $k$), the biverse spectrum starts
oscillating (see also Fig.~\ref{fig:deltaTau_hk}). The power spectrum is taken at the time $\eta^*$ such that $|V_{\text{eff}}(\eta^*)| = k_{\mathrm{max}}^2$.
}
\label{fig:deltaTau_Pk}
\end{figure}

Finally, Fig.~\ref{fig:deltaTau_Pk} shows the tensor power spectrum resulting from the single state trajectory and the biverse case. 
As in the previous case, the biverse spectrum has almost the same spectral index as the single state case for small $k-$modes, but shows stark deviations for wavenumbers above $k\sim 4\times 10^{-3}$, after
which the spectrum starts oscillating. Any effect of the
superposition of states enters only for the larger part of the chosen
wavenumber spectrum, and thus for relatively smaller
wavelengths.

The question of observability (or reverse engineering) of trajectories
in a multi-state configuration boils down to determining the actual
range of measurable scales involved in our calculation. As already
mentioned above, this can only be addressed within a full cosmological
model describing the total evolution of the Universe from the
bounce to now. Our model, only focusing on the quantum
bounce, does not allow such a connection with present-day observations, 
but we can make first statements about possible effects:
Supposing
for instance that the relevant range ends where the spectrum starts
bending in 
Fig.~\ref{fig:deltaTau_Pk}, fitting the spectrum to a single power law would result in
a smaller spectral index w.r.t. the single state case. 



\section{Conclusions}
\label{sec:conc}

We have presented a canonical quantization based (WDW equation)
quantum cosmological model for which the minisuperspace for the
evolution of the background state is supplemented with tensor
perturbations. The scale factor being a non-negative definite
variable, we use a scheme suitable for the half-line, namely affine
quantization using coherent states. This leads to the resolution of
the Big-Bang singularity which is replaced by a bounce.

To evaluate the wave function of the perturbed universe, one usually
assumes a Born-Oppenheimer approximation, in which the full state can
be written as a direct product of background and perturbations, to
hold at all times. Going beyond this approximation is known to produce
potentially observable consequences in the form of non-trivial
non-gaussianities even with vacuum initial fluctuations stemming from
a purely gaussian theory~\cite{Bergeron:2025eda}. It is usually not
emphasized that this procedure requires that the expectation value of
the background state can consistently be interpreted as giving an
effective evolution of the quantum corrected scale factor, which then
enables one to understand the impact of quantum effects on the
evolution of perturbations.  The effective scale factor, which usually
resolves the singularity, is then treated simply as a semi-classical
source to produce particles, which in turn produce primordial
fluctuations that subsequently evolve into large scale structures.  In
contrast to previous works, we considered here quantum trajectories as
an alternative approach to obtaining the background evolution of the
universe on which perturbations propagate, thus proposing a more
general procedure that should apply even in cases in which the
standard does not.

For the sake of simplicity, we considered a model for which the
Universe is dominated by a single perfect fluid, which also serves as
a clock.  In practice, the fluid can be identified with radiation, as
it can be argued that quantum gravity effects should become relevant
only in very early epochs during which the curvature is sufficiently
high and during which one can reasonably expect all the matter content
to behave relativistically.  For an FLRW universe filled with a
perfect fluid, a basis of coherent states describing the background
Hilbert space is known~\cite{Bergeron:2023zzo}, where each of these
states represents a good candidate for  the description of the
background through a suitable semi-classical state; they all follow
well-defined bouncing trajectories that can be obtained from a
semi-classical Hamiltonian and are parametrized by their definite
energies and bounce times.

Using this basis, one can consider a more general state as any linear
superposition of such states with different energies $E$ and bouncing
times $\tau_\textsc{b}$, weighed by an arbitrary function
$W(E,\tau_\textsc{b})$. Such states, while unthinkable in a classical
context, are perfectly acceptable in the quantum regime.  It turns out
that provided the weight function is of compact support in both
variables, the asymptotic behavior is always well approximated by a
single state and thus agrees with a single peaked semi-classical
trajectory far from the full quantum regime, retrieving the dynamics
of GR.

In the full quantum regime however, it is not evident how to define
the effective trajectories of the scale factor and thus the background
from the superposition state unless one assumes a formulation of
quantum physics that includes them naturally through the eikonal
approximation.  Such trajectories are unambiguous (see however
Ref.~\cite{Deotto:1997ci}) and permit to solve several issues related
with the measurement problem at the price of loosing locality; they
can also be subject to interpretational questions in some
context~\cite{Englert1992, Naaman_Marom_2012, Hiley:2018ofx,
Tastevin_2018}.

Still assuming the Born-Oppenheimer approximation, one then presumes
the background state to solve the zeroth order time-dependent
Schr\"odinger equation, which describes the background evolution, each
solution of which leading to a congruence of trajectories equivalent
to the usual probability flow.  The actual Universe then follows one
and only one such trajectory, i.e., given a state that provides an
infinite number of possible trajectories, only one of these
trajectories is selected by fixing an initial condition, even with
non-classical superpositions.  This trajectory gives the unambiguous
evolution of the background (i.e., the scale factor) even in the full
quantum regime.

Once a trajectory has been picked, the next step of the method to
compute perturbations consists in replacing occurrences of the
background quantities by their actual time-dependent functions as
determined by the quantum trajectories in the second-order Hamiltonian
that describes the evolution of perturbations. From that point on, the
standard treatment for cosmological perturbation
theory~\cite{Mukhanov:1990me} can be straightforwardly applied and
$n-$point functions calculated.

In the single state background, the above-defined trajectories
are proportional to the semi-classical ones resulting from taking
the expectation value, so the ensuing impact on the perturbation
spectrum and properties is unchanged whether or not such trajectories
are given a physical reality.
A superposed background
state however, while equivalent to the single state case and GR in the
semi-classical regime, may differ tremendously in the quantum regime,
opening up a large variety of possibilities of consequences on
perturbations and can thereby in principle be constrained by
observations.

Comparing the situation examined in the present work with that of
Ref.~\cite{Bergeron:2025eda}, it is tempting to consider such a simple
cosmological model to address the issue of the meaning and
consequences of eikonal trajectories in quantum physics: starting with
the same initial conditions and ending with the same background
configuration, the two descriptions are expected to yield potentially
distinct predictions for the power spectra of perturbations, some
originating as direct consequences of the existence of trajectories.
Importantly, while the details of the novel features introduced by the
trajectories are highly sensitive to the chosen parameters and initial
conditions, their occurrence and deviation from the standard treatment
is not.

In order to establish the possibility of such observable imprints, we consider the time development of tensor perturbations in a universe
dominated by a radiation fluid. Although non-realistic because
incomplete, this example permits to exhibit interesting properties.
We thus calculate explicitly the tensor power spectrum for two
specific examples and compare the resulting spectra to those one
would obtain in the single state case.

The dynamics of the perturbation modes are found to be
influenced by the superposition trajectories, but in a way
that may or may not be measurable.  
In particular, we found
that in the limit of 
small wavenumbers, the
spectral index $n_\textsc{t}$ 
takes on
its one-bounce value
$n_\textsc{t} = 2$ in the cases we focused on, with only a
slight modification (decrease) of its amplitude. For larger wavenumbers,
and therefore smaller scales, the spectrum 
highly depends on the concrete realization of the biverse. As one
would expect, if the trajectory is not very different from that
of a single universe, 
only the amplitude changes, 
and thus no effect could be measured in 
cosmological
observations. In a more drastic case, i.e. when the biverse
trajectory displays a second bounce, the power spectrum 
can exhibit non-linear features that alter the spectral index
(if one insists on fitting a simple power law).

In a forthcoming work, we plan to study how such
changes may concretely manifest in
observations and what generic predictions could be made more
explicitly, including the scalar modes.
At this stage, we would like to emphasize that the existence
of trajectories for superposition states \emph{can} lead to 
measurable effects, as examplified by the change in the
primordial stochastic gravitational
wave (tensor mode) background.
Furthermore, it would be interesting to
establish how going beyond the Born-Oppenheimer approximation in the
trajectory approach would affect these predictions. 
Paraphrasing
EPR~\cite{Einstein:1935rr}, one could then ask the question "Can
quantum-mechanical description of cosmology be considered complete?",
a question that could perhaps lead to observational proposals and
yield potentially new insights into the quantum realm in general.

\begin{widetext}
\appendix*
\section{Practical quantization map}

The quantization map \eqref{quantMap} provides a natural way
to transform any function of phase space $f(q,p)$ into an operator
$\hat{A}_f$ acting on the Hilbert space $\mathscr{H}$. 
While already used in several previous works, we illustrate this quantization procedure below for the convenience of the reader.
In practice,
one starts from the ``$x$'' representation of the coherent state \eqref{eq:cohState},
namely
\begin{align}
\braket{x}{q,p}_{\psi_0} = \bra{x} e^{\ii \frac{p}{2q} \hat{X}^2}
e^{-\frac12 \ii \ln q \qty( \hat{X} \hat{P} +
\hat{P} \hat{X})}\ket{\psi_0}
= \dfrac{1}{\sqrt{q}} e^{\ii \frac{p}{2q} x^2}
\psi_0\qty(\dfrac{x}{q}),
\end{align}
where $\ket{\psi_0}$ denotes the fiducial state vector and we used, still in the ``$x$'' representation, the fact that
\begin{align}
\bra{x}\frac12 \qty( \hat{X} \hat{P} + \hat{P} \hat{X})\ket{\psi_0}
= -\ii \qty(\dfrac{\dd}{\dd \ln x} +\dfrac12 ) \psi_0(x),
\end{align}
 so that 
$$
\bra{x} e^{-\frac12 \ii \ln q \qty( \hat{X} \hat{P} +
\hat{P} \hat{X})}\ket{\psi_0} = \dfrac{1}{\sqrt{q}}
\psi_0\qty(\dfrac{x}{q}).
$$
One can then calculate the ``$x$'' representation of any operator 
corresponding to a phase space function $f(q, p)$ as  \eqref{quantMap}
\begin{align}
\bra{x} \hat{A}_f (\psi_0) \ket{\varphi} = \mathcal{N}_{\psi_0} 
\int_{\mathbb{R} \times \mathbb{R}^+}
\dd{q}\dd{p} \braket{x}{q,p}_{\psi_0}\, f(q,p) \tensor[_{\psi_0}]{\braket{q,p}{\varphi}}{}\, ,
\label{quantMap1}
\end{align}
which, for $f(q,p)=1$ leads to the closure relation
\begin{align}
\bra{x} \hat{A}_1 (\psi_0) \ket{\varphi} & = \mathcal{N}_{\psi_0} 
\int_{\mathbb{R} \times \mathbb{R}^+}
\dd{q}\dd{p} \braket{x}{q,p}_{\psi_0}   
\tensor[_{\psi_0}]{\braket{q,p}{\varphi}}{}\\
& = \mathcal{N}_{\psi_0} 
\int_{\mathbb{R} \times \mathbb{R}^+}
\dd{q}\dd{p} \braket{x}{q,p}_{\psi_0}\int_{\mathbb{R}^+} \dd{y}
\tensor[_{\psi_0}]{\braket{q,p}{y}}{} \braket{y}{\varphi} \\
& = \mathcal{N}_{\psi_0} 
\int_{\mathbb{R} \times \mathbb{R}^+}
\dd{q}\dd{p} 
\dfrac{1}{\sqrt{q}} e^{\ii \frac{p}{2q} x^2}
\psi_0\qty(\dfrac{x}{q})
\int_{\mathbb{R}^+} \dd{y}
\dfrac{1}{\sqrt{q}} e^{-\ii \frac{p}{2q} y^2}
\psi_0\qty(\dfrac{y}{q}) \varphi(y),
\end{align}
where we consider a real fiducial function $\psi_0 (x)\in \mathbb{R}$.

The canonical change of variable $Q=q^2$ and $P=p/(2q)$ then permits
to simplify the integrals, and using
\begin{align}
\int_{\mathbb{R}} \dd{P} e^{-\ii P (y^2-x^2)} = \dfrac{\pi}{x} \delta(x-y)\,,
\end{align}
one gets
\begin{align}
\bra{x} \hat{A}_1 (\psi_0) \ket{\varphi} = \pi \mathcal{N}_{\psi_0}
\int_{\mathbb{R}^+} 
\dfrac{\dd{Q}}{\sqrt{Q}}  \frac{\varphi(x)}{x} \psi_0^2 \qty(
\dfrac{x}{\sqrt{Q}}),
\end{align}
and finally, upon setting $z=x/\sqrt{Q}$, 
\begin{align}
\bra{x} \hat{A}_1 (\psi_0) \ket{\varphi} = \qty[ 2 \pi \mathcal{N}_{\psi_0}
\int_0^\infty \dfrac{\dd{z}}{z^2}\psi_0^2 \qty(z) ] \, 
\varphi(x).
\end{align}
The relation $\bra{x} \hat{A}_1 (\psi_0) \ket{\varphi} =
\varphi(x)$ thus requires the normalization constant
to be
\begin{align}
\mathcal{N}^{-1}_{\psi_0} = 
2 \pi \int_0^\infty \dfrac{\dd{z}}{z^2}\psi_0^2 \qty(z) = 2\pi
c_0(\psi_0),
\end{align}
where we defined the generic integrals
\begin{align}
c_\gamma (\psi_0) = \int_0^\infty \dfrac{\dd{z}}{z^{2+\gamma}}
\psi_0^2 \qty(z).
\end{align}
Similar calculations can be done for other operators,
leading to
\begin{align}
\bra{x} \hat{A}_q (\psi_0) \ket{\varphi} = \qty[
\dfrac{c_1(\psi_0)}{c_0(\psi_0)}] \, x \varphi(x),
\end{align}
and
\begin{align}
\bra{x} \hat{A}_p (\psi_0) \ket{\varphi} = \qty[
\dfrac{c_1(\psi_0)}{c_0(\psi_0)}]
\, \qty[ -\ii \dfrac{\dd}{\dd x} \varphi(x)],
\end{align}
so that imposing the normalization $c_1(\psi_0) = c_0(\psi_0)$ for the
fiducial state $\psi_0(x)$ allows to recover the usual Dirac algebra
$\comm{\hat{A}_q}{\hat{A}_p}=\ii$.

Calculating the Hamiltonian map $\hat{A}_{p^2}$ is slightly more
involved and requires some additional integrations by part, and,
although tedious, is rather straightforward. 
 Further details on the explicit calculation to obtain $\hat{A}_{p^2}$ can be found in e.g. \cite{Bergeron:2013ika}.
One finally gets
\begin{align}
\bra{x} \hat{A}_{p^2} (\psi_0) \ket{\varphi} = \qty[
\dfrac{c_2(\psi_0)}{c_0(\psi_0)}]
\, \underbrace{\qty[ - \dfrac{\dd^2}{\dd x^2} 
\varphi(x)]}_{=\bra{x}\hat{P}^2\ket{\varphi}} + 
\dfrac{K(\psi_0)}{c_0(\psi_0)} \underbrace{\dfrac{\varphi(x)}{x^2}}
_{= \bra{x}\hat{X}^{-2}\ket{\varphi}},
\label{AppP2}
\end{align}
where  $K(\psi_0) = \int \dd y\, y^{-2} \psi_0(y)^2 - \frac{3}{2} c_2$ is an
integral of $\psi_0$, which can be shown to be non-negative~\cite{Bergeron:2023zzo}. 
Upon rescaling time in the Schr\"odinger equation $\hat{\mH}_T\psi = (\hat{A}_{p^2} - \ii \pd_\tau )\psi =0 $ as $\tau \to \frac{c_2}{c_0}\tau$, one can get
rid of the first pre-factor $c_2/c_0$, yielding the form
\eqref{Hquant} as announced.
The comparison to \eqref{Hquant} then reveals that $\nu^2 = K(\psi_0)/c_2(\psi_0) + \frac{1}{4}$ and is thereby fully determined by the choice of $\psi_0$. The interested reader is referred to 
Ref.~\cite{Bergeron:2023zzo} or more details.

\end{widetext}

\acknowledgments
The work of LM is funded by the Leverhulme Trust through a Study
Abroad Studentship.

\bibliography{Refs.bib}

\end{document}